\documentclass[journal]{IEEEtran}

\usepackage{array}
\usepackage[dvipdfmx]{graphicx} 
\graphicspath{{images/}}
\DeclareGraphicsExtensions{.pdf,.jpg,.png}
\usepackage[fleqn]{amsmath}
\usepackage{algorithm}
\usepackage{algorithmicx}
\usepackage{algpseudocode}
\ifCLASSOPTIONcompsoc
 \usepackage[caption=false,font=normalsize,labelfont=sf,textfont=sf]{subfig}
\else
 \usepackage[caption=false,font=footnotesize]{subfig}
\fi
\usepackage{fixltx2e}

\usepackage{cite}
\usepackage[acronym]{glossaries}
\usepackage{mathtools}
\usepackage{amsfonts}
\usepackage{amssymb}
\usepackage{bbold} 
\usepackage{bm} 
\usepackage[dvipdfmx]{pdfcomment} 
\usepackage{multirow}
\usepackage{booktabs} 
\usepackage{threeparttable} 
\usepackage{color} 
\usepackage{grffile}
\usepackage{xspace}

\newtheorem{assumption}{Assumption}
\newtheorem{theorem}{Theorem}
\newtheorem{lemma}{Lemma}

\newenvironment{revhl}[1]{}{}
\newcommand{\revhlnon}[1]{#1}
\newcommand{\revhlm}[2]{#2}

\def\figsizeConcept{0.49\textwidth}
\def\figsizeToymodel{0.30\textwidth}
\def\figsizeDistillation{0.45\textwidth}
\def\figsizeTopology{0.09\textwidth}
\def\figsizeConvergenceAll{0.49\textwidth}
\def\figsizeRobustAll{0.49\textwidth}
\def\figsizeMixedAll{0.49\textwidth}
\def\figsizeMixedThree{0.40\textwidth}
\def\figsizeManyAgentsAll{0.40\textwidth}

\newacronym{IID}{IID}{independent and identically distributed}
\newacronym[plural={IoE}]{IoE}{IoE}{Internet of Everything}
\newglossaryentry{Frechet}{name={Fr\'echet}, description={}}
\newacronym{FL}{FL}{federated learning}
\newacronym{DL}{DL}{deep learning}
\newacronym{NN}{NN}{neural network}
\newacronym{DNN}{DNN}{deep neural network}
\newacronym{ML}{ML}{machine learning}
\newacronym{GBDT}{GBDT}{gradient boosting decision tree}
\newacronym{MSE}{MSE}{mean squared error}
\newacronym{KL}{KL}{Kullback-Leibler}
\newacronym{CDO}{CDO}{consensus-based distributed optimization}
\newacronym{CMFD}{CMFD}{consensus-based multi-hop federated distillation}
\newacronym{WSN}{WSN}{wireless sensor networks}
\newacronym{P2P}{P2P}{peer-to-peer}
\newacronym{BA}{BA}{Barab\'{a}si--Albert}
\newacronym{SGD}{SGD}{stochastic gradient descent}
\newacronym{DPSGD}{D-PSGD}{decentralized parallel \gls{SGD}}
\newacronym{FMNIST}{F-MNIST}{fashion MNIST}
\newacronym{GAN}{GAN}{generative adversarial network}
\newacronym{FD}{FD}{federated distillation}
\newacronym{FAug}{FAug}{federated augmentation}
\newacronym{RHS}{RHS}{right-hand side}
\newacronym{RKHS}{RKHS}{reproducing kernel {Hilbert} space}
\newacronym{DGD}{DGD}{distributed gradient descent}
\newacronym{ADMM}{ADMM}{alternating direction method of multipliers}
\newacronym{GADMM}{GADMM}{group \gls{ADMM}}
\newacronym{D-ADMM}{D-ADMM}{distributed \gls{ADMM}}
\newacronym{PDMM}{PDMM}{primal-dual method of multipliers}

\newcommand{\etal}{\textit{et al.\ }}

\renewcommand{\[}{\left[}
\renewcommand{\]}{\right]}
\renewcommand{\(}{\left(}
\renewcommand{\)}{\right)}
\newcommand{\relmiddle}[1]{\mathrel{}\middle#1\mathrel{}}
\renewcommand{\mid}{\relmiddle|}
\newcommand{\lv}{\left\lVert}
\newcommand{\rv}{\right\rVert}
\newcommand{\Dist}[1]{\lv #1 \rv}
\newcommand{\Real}{\mathbb{R}}

\newcommand{\Neigh}[1]{\mathcal{N}_{#1}}
\newcommand{\Laplacian}{L}
\newcommand{\NumNeigh}{n}

\newcommand{\sharedD}{\mathcal{D}_\mathrm{s}}
\newcommand{\Devices}{\mathcal{U}}
\newcommand{\Ds}{\mathcal{D}}
\newcommand{\Dsi}{\mathcal{D}_i}
\newcommand{\Dsm}{\hat{\mathcal{D}}}

\DeclareMathOperator*{\minimize}{minimize}
\DeclareMathOperator*{\argmin}{arg\,min}
\DeclareMathOperator*{\argmax}{arg\,max}

\newcommand{\Lmu}{\mathsf{L}_{\mu}}
\newcommand{\Lmui}{\mathsf{L}_{\mu_i}}
\newcommand{\dmu}{\mathrm{d}\mu}
\newcommand{\dmui}{\mathrm{d}\mu_i}
\newcommand{\InSpc}{\mathcal{X}}
\newcommand{\OutSpc}{\mathcal{Y}}
\newcommand{\FnSpc}{\mathcal{F}}
\newcommand{\LSpc}[1][\mu]{L^2_{#1}}
\newcommand{\subDifLmui}[1][f_i^t]{\partial\Lmui(#1)}
\newcommand{\FedSpc}[1][n]{\Phi_{#1}}
\newcommand{\Users}{\mathcal{U}}
\newcommand{\DimX}{N}
\newcommand{\DimY}{M}
\newcommand{\NumU}{n}
\newcommand{\Lip}[1]{K_{#1}}
\newcommand{\maxRn}{S_i}
\newcommand{\maxLip}{\Lip{\mathrm{m}}}
\newcommand{\trueF}{f^\star}
\newcommand{\rn}{\nu}
\newcommand{\inner}[3]{\langle #1, #2 \rangle_{#3}}
\newcommand{\lr}{\eta}
\newcommand{\lrc}{\varepsilon}

\newcommand{\ws}[1][i]{{\hat{\bm{w}}_{#1}}}
\newcommand{\predy}[1]{{\hat{\bm{y}}_{#1}}}
\newcommand{\Trans}{\mathrm{T}}
\newcommand{\Ones}[1][\NumU\times\NumU]{\bm{1}_{#1}}
\newcommand{\FedMeanF}{\bar{\bm{f}}}
\newcommand{\FedMeanD}{\bar{\bm{d}}}
\newcommand{\MeanF}{\bar{f}}
\newcommand{\MeanD}{\bar{d}}
\newcommand{\bestF}{f_{\mathrm{best},t}}
\newcommand{\grad}[2]{d_{#1}^{#2}}
\newcommand{\gradAvg}{d_{\MeanF_t}}
\newcommand{\MaxD}{\Delta}
\newcommand{\DimW}{n_w}
\newcommand{\Card}[1]{\vert{#1}\vert}
\newcommand{\eigen}{\lambda}
\newcommand{\eigenQ}{\kappa}
\newcommand{\convRate}{\kappa_2}
\newcommand{\distAvg}{\gamma}
\newcommand{\initF}{\Dist{\bm{f}_1}_{\FedSpc}}
\newcommand{\initDist}{C_1}
\newcommand{\Coeff}{C_2}
\newcommand{\MeanDist}{D}

\begin{document}
\begin{figure*}
\Large{\copyright\ 2022 IEEE. Personal use of this material is permitted. Permission
from IEEE must be obtained for all other uses, in any current or future
media, including reprinting/republishing this material for advertising or
promotional purposes, creating new collective works, for resale or
redistribution to servers or lists, or reuse of any copyrighted
component of this work in other works.}
\end{figure*}
\clearpage
\setcounter{page}{1}
\title{Decentralized and Model-Free Federated Learning: Consensus-Based Distillation in Function Space}
%
%

\author{Akihito~Taya,~\IEEEmembership{Member,~IEEE,}
        Takayuki~Nishio,~\IEEEmembership{Senior~Member,~IEEE,}
        Masahiro~Morikura,~\IEEEmembership{Member,~IEEE,}
        and~Koji~Yamamoto,~\IEEEmembership{Senior~Member,~IEEE}
\thanks{Akihito Taya
is with the College of Science and Engineering, Department of Integrated Information Technology, Aoyama Gakuin University, Sagamihara 252-5258, Japan
(e-mail: taya@it.aoyama.ac.jp)}
\thanks{Takayuki Nishio
is with the School of Engineering, Tokyo Tech., Japan
(e-mail: nishio@ict.e.titech.ac.jp)}
\thanks{Masahiro Morikura and Koji Yamamoto
are with the Graduate School of Informatics, Kyoto University, Kyoto 606-8501, Japan
(e-mail: kyamamot@i.kyoto-u.ac.jp}%
\thanks{Manuscript received April 19, 2005; revised August 26, 2015.}}

%
%

\markboth{Journal of Sig..,~Vol.~XX, No.~XX, January~2020}%
{Shell \MakeLowercase{\textit{et al.}}: Bare Demo of IEEEtran.cls for IEEE Journals}
%


\maketitle

\begin{abstract}
This paper proposes a fully decentralized \gls{FL} scheme for \gls{IoE} devices that are connected via multi-hop networks.
Because \gls{FL} algorithms hardly converge the parameters of \gls{ML} models,
this paper focuses on the convergence of \gls{ML} models in \textit{function spaces}.
Considering that the representative loss functions of \gls{ML} tasks e.g., \gls{MSE} and \gls{KL} divergence, are convex \textit{functionals},
algorithms that directly update functions in function spaces could converge to the optimal solution.
The key concept of this paper is
to tailor a consensus-based optimization algorithm to work in the function space
and achieve the global optimum in a distributed manner.
This paper first analyzes the convergence of the proposed algorithm in a function space, which is referred to as a meta-algorithm,
and shows that the spectral graph theory can be applied to the function space in a manner similar to that of numerical vectors.
Then, \gls{CMFD} is developed for a \gls{NN} to implement the meta-algorithm. 
\Gls{CMFD} leverages knowledge distillation to realize function aggregation among adjacent devices without parameter averaging.
An advantage of \gls{CMFD} is that it works even with different \gls{NN} models among the distributed learners.
Although \gls{CMFD} does not perfectly reflect the behavior of the meta-algorithm,
the discussion of the meta-algorithm's convergence property promotes an intuitive understanding of \gls{CMFD},
and simulation evaluations show that \gls{NN} models converge using \gls{CMFD} for several tasks.
The simulation results also show that \gls{CMFD} achieves higher accuracy than parameter aggregation for weakly connected networks,
and \gls{CMFD} is more stable than parameter aggregation methods.
\end{abstract}


%
\IEEEpeerreviewmaketitle

\glsresetall

\section{Introduction} \label{sec:intro}
\IEEEPARstart{T}{he} \gls{IoE}
is a paradigm in which commonly used devices (e.g., machines, smartphones, cars, houses, and also humans)
are connected and interact with each other.
These devices have various sensors to perceive the real world and share information.
Such data accelerate data-driven analysis and \gls{ML}, particularly \gls{DL}.
\Gls{IoE} devices are able to provide \gls{ML} agents with images, sound, physiological signals, and various data to train \gls{ML} models
for semantic analysis, prediction, as well as many other tasks.
However, several challenges are encountered when utilizing the sensor data for \gls{ML}.
Protecting privacy is one of the most important problems.
For example, it is preferable not to upload users’ photos, conversations, and behavioral data stored in \gls{IoE} devices to data centers.

\begin{figure}[!t]
\centering
\includegraphics[width=\figsizeConcept]{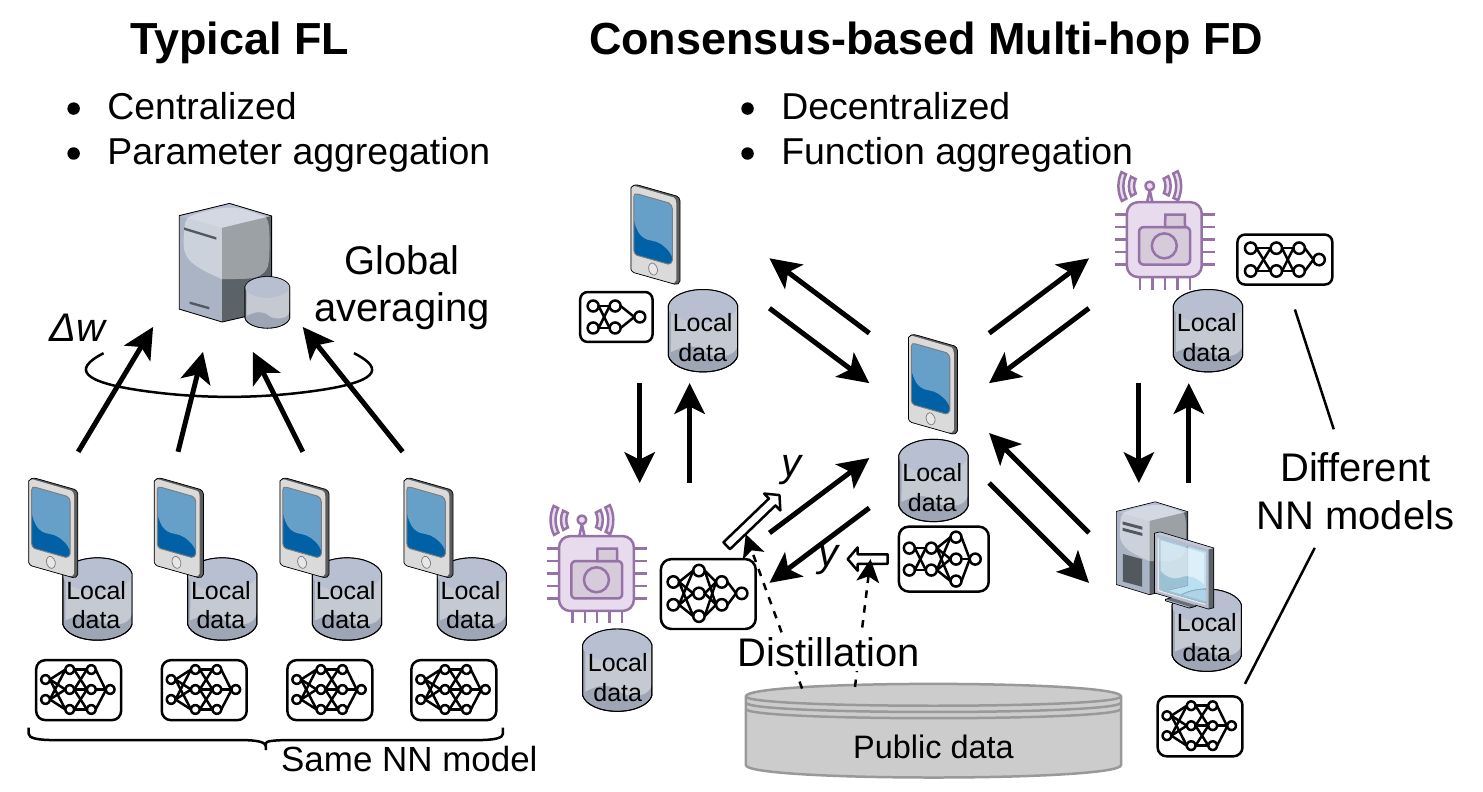}
\caption{
Concept of \gls{CMFD}.
In typical \gls{FL} schemes \cite{mcmahan2016communication}, devices send parameters to the server, where the parameters are aggregated by averaging methods.
With \gls{CMFD}, the devices share output values with the adjacent devices via multi-hop networks
and update their local models by distillation.
\Gls{CMFD} realizes scalable and model-free \gls{FL} systems.
}
\label{fig:concept}
\end{figure}

\Gls{FL} has been proposed to avoid uploading data for preserving user privacy \cite{mcmahan2016communication,kairouz2019advances}.
Instead of collecting data to a server,
\Gls{FL} updates \gls{NN} models at local devices (e.g., smartphones) using their private data
and uploads the updated \gls{NN}-parameters (i.e., weights and biases of each layer) to the server.
Then, the server gathers and aggregates the uploaded parameters
and broadcasts the up-to-date parameters to the local devices.
The left side of Fig.~\ref{fig:concept} shows the parameter-aggregation scheme.
This algorithm realizes distributed learning without exposing the user data to the centric learner.
Since the \gls{FL} was first introduced in \cite{mcmahan2016communication},
many types of \gls{FL} have been developed to solve problems that appear in practical situations \cite{kairouz2019advances}.

\glsreset{CMFD} 

\begin{table*}[!t]
\centering
\begin{threeparttable}
\caption{Algorithm comparison} \label{tbl:comparison}
\begin{tabular}{ccccccc}
\toprule
& Network topology & Algorithm & Aggregation method & Model-free & IIDness & Convergence analysis \\
\midrule
FedAvg \cite{mcmahan2016communication}
  & Star topology & Centralized & Parameter averaging & No & non-IID & N/A \\
\cite{savazzi2020federated}
  & Multi-hop & Decentralized & Parameter averaging & No & non-IID & N/A \\
\cite{lalitha2019peer,lian2017can,niwa2020edge}
  & Multi-hop & Decentralized & Parameter averaging & No & non-IID & parameter convergence \\
\cite{sato2020network}
  & Multi-hop & Decentralized & Parameter averaging & No & IID & N/A \\
\cite{oh2020mix2fld}
  & Star topology & Centralized & Distillation & No & non-IID & N/A \\
\cite{ahn2019wireless,itahara2021distillation,jeong2018communication}
  & Star topology & Centralized & Distillation & No\tnote{a} & non-IID & N/A \\
\cite{chang2019cronus,li2019fedmd,lin2020ensemble}
  & Star topology & Centralized & Distillation & Yes & non-IID & N/A \\
\cite{jeong2019multi}
  & Multi-hop & Centralized & Distillation & No & non-IID & N/A \\
\cite{anil2018large,zhang2018deep}
  & Complete graph & Decentralized & Distillation & No\tnote{a} & IID & N/A \\
CMFD (ours)
  & Multi-hop & Decentralized & Distillation & Yes & non-IID & function convergence \\
\bottomrule
\end{tabular}
\begin{tablenotes}
  \item[a] It may be possible to implement, but was not evaluated.
\end{tablenotes}
\end{threeparttable}
\end{table*}

Typical \gls{FL} schemes require servers to manage client devices and aggregate parameters,
whereas \gls{IoE} devices are expected to operate without a server because they are sometimes connected via multi-hop networks,
through which devices can directly communicate only with adjacent ones.
This constraint makes it difficult to adopt \gls{FL}, which requires a central server \cite{savazzi2020federated}.
Therefore, multi-hop \gls{FL} algorithms have been developed \cite{lalitha2019peer,savazzi2020federated,lian2017can,niwa2020edge,sato2020network},
through which devices share information only with adjacent devices.
Multi-hop \gls{FL} can solve some problems of centralized \gls{FL}.
First, multi-hop \gls{FL} does not suffer from communication bottlenecks in the servers.
Therefore, the number of \gls{FL}-participating devices is easily increased,
and a large amount of training data can be utilized for learning.
Second, considering that some information is sent to other devices even if it does not include private data,
users often feel concerns when sending information to unknown devices.
Therefore, the study of multi-hop \gls{FL} should be motivated to realize higher scalability and security than a centralized one.

Although multi-hop \gls{FL} algorithms were studied in \cite{lalitha2019peer,savazzi2020federated,lian2017can,niwa2020edge,sato2020network},
the following challenges remain.

\subsubsection*{Model constraint}
Existing works assume that all the devices have the same \gls{NN} model architecture
(e.g., the number of layers, hidden nodes, and how they are connected).
However, considering that \gls{IoE} devices have different computational resources,
it is better to use different \gls{NN} models depending on their computational capacities.
Besides, if some \gls{IoE} device vendors collaborate in a single \gls{FL} system,
they may prefer to keep their \gls{NN} models secret.
Therefore, the constraint should preferably be eased
and allow the devices to decide their \gls{NN} models by themselves depending on their own conditions.

\subsubsection*{Convergence analysis under non-IID situations}
The convergence of multi-hop \gls{FL} is not guaranteed
because multi-hop \gls{FL} tasks are formulated as decentralized non-convex optimization problems.
If centralized \gls{FL} algorithms are adopted,
all devices can synchronize their parameters at the end of each iteration
because there is a parameter server that manages and broadcasts a global model.
Conversely, it is not generally guaranteed when using decentralized algorithms in which all devices obtain the same model
because there is no opportunity for synchronizing all models.
In addition, multi-hop \gls{FL} more severely suffers from non-\gls{IID} problems than centralized \gls{FL}.
It is usual in \gls{FL} that data distributions vary across devices because of users' interests and conditions \cite{kairouz2019advances}.
Such situations, termed non-\gls{IID}ness, which has been widely studied \cite{zhao2018federated,sattler2020robust,li2020convergence},
drastically degrades \gls{FL} performance.
In contrast to centralized \gls{FL},
multi-hop \gls{FL} is further affected by non-\gls{IID}ness because \gls{NN} models are shared only among adjacent devices.

To solve these problems, we first propose a decentralized \gls{FL} algorithm
that optimizes the prediction functions in a \textit{function space}.
The proposed algorithm is referred to as a meta-algorithm in this paper
because it can be applied to \glspl{NN} and other gradient-descent-based \gls{ML} algorithms, such as the \gls{GBDT}.
The meta-algorithm iteratively trains the local prediction functions using functional gradients
and aggregates them with adjacent devices by applying \gls{CDO} to guarantee convergence even in multi-hop networks.
Because typical \gls{ML} tasks are convex functional optimization problems,
the meta-algorithm, which works in a function space and does not handle parameters,
can solve them, and convergence is achieved by gradient-descent-based methods using functional gradients.

\begin{revhl}{1-1}
To improve the practicality, we adopt a gradient descent scheme, which is widely spread and easily implementable.
In the context of \gls{IoE}, where different manufacturers develop diverse devices,
some devices do not have enough computational resources to implement complicated algorithms,
and some manufacturers prefer to reduce the development costs.
\end{revhl}

The convergence of the meta-algorithm is mathematically analyzed in a function space
because the convergence in a function space is practically sufficient,
whereas the convergence in a parameter space is difficult to achieve.
To analyze how local prediction functions draw close in a function space when using the meta-algorithm,
we utilize spectral graph theory to prove convergence.
The analysis is divided into two parts:
the convergence among the devices provided in Theorem~\ref{th:convdist}
and the convergence to the global optimum provided in Theorem~\ref{th:convopt}.
\begin{revhl}{1-1}
Through the analysis, we do not use the information of the representation of \gls{ML} models
considering the situations where some types of models are adopted by different devices
and they do not share their information.
We develop an analysis scheme that utilizes only function distances
because the distance can be calculated without specifying the representation.
\end{revhl} 

\begin{revhl}{1-1}
By focusing on the behavior of gradient descent in a function space,
we successfully proved the convergence of the meta-algorithm and provided an upper bound under non-\gls{IID} situations.
We also revealed the effects of network connectivity on the convergence rate in the function space.
To do so, we proved Lemma~\ref{lemm:inducednorm}, which is used in the proof of Theorem~\ref{th:convdist}, to analyze the convergence of functions.
Lemma~\ref{lemm:inducednorm} states that the induced norm of matrices corresponding in real-vector and stacked function spaces are the same.
This fact enables us to analyze convergence in function spaces in a similar manner as that in numerical vector spaces.
\end{revhl}
It is also notable that the meta-algorithm operates in a decentralized manner and is scalable to the number of devices owing to the \gls{CDO} scheme.

A \gls{CMFD} is developed as a \gls{NN} version of the meta-algorithm because it cannot be implemented directly.
Fig.~\ref{fig:concept} shows an overview of our \gls{CMFD},
and the characteristics of \gls{CMFD} are compared with other \gls{FL} algorithms in Table~\ref{tbl:comparison}.
\gls{CMFD} addresses one of the impracticalities of the meta-algorithm: the difficulty of aggregating prediction functions in a function space.
Because boosting algorithms, for example, \gls{GBDT}, are functional gradient methods as explained in \cite{mason1999boosting},
the meta-algorithm is easily applied to boosting algorithms by considering weak learners as functional gradients.
However, \gls{SGD}-based algorithms for \gls{NN} are parametric, and an implementation scheme must be developed.
The calculation of aggregated prediction functions among adjacent devices is a technical problem
because it is difficult to determine the parameters $\bm{w}$ that satisfy $f(\bm{x};\bm{w}) = \sum_i f_i(\bm{x})$ for given prediction functions $f_i$.
\begin{revhl}{1-4}
We show a toy model how \gls{CMFD} tackles function aggregation using distillation in Appendix \ref{sec:app_toymodel}.
\end{revhl}

Function aggregation is not a challenge when adopting ensemble \gls{ML} or kernel machines \cite{koppel2018decentralized},
because the aggregated functions can be calculated as the linear sum of the weak learners or kernel functions.

We utilize distillation schemes \cite{bucilua2006model} to adjust the parameters of \gls{NN}s to achieve aggregation in function spaces.
In the \gls{CMFD}, the outputs of local functions of adjacent devices are aggregated,
and the devices train their local functions regarding the aggregated outputs as labels.
The local functions approach the aggregated function using this scheme.
Although distillation requires the same inputs to be shared, public open data can be utilized to avoid sharing private data \cite{itahara2021distillation}.
In addition, distillation allows the devices to adopt different \gls{NN} models
because the devices do not directly share their parameters.
Thus, the devices can independently update the parameters of their \gls{NN} models without considering the parameters of other devices.
This scheme relaxes the constraints of \gls{FL}
and is suitable for \gls{IoE} systems, where the devices have different computational resources.
Distillation methods are applied in centralized \gls{FL} \cite{park2019wireless,park2019distilling,oh2020mix2fld,ahn2019wireless,itahara2021distillation,chang2019cronus,jeong2018communication}
because of communication efficiency.
However, to the best of our knowledge, distillation-based \gls{FL} in multi-hop networks has not been studied.

The contributions of this paper are summarized as follows:
\begin{enumerate}
\item
To decentralize the \gls{ML} models' training for \gls{IoE} systems,
we propose an algorithm, termed a meta-algorithm, that optimizes \gls{ML} models in a \textit{function space}.
The meta-algorithm adopts a \gls{CDO} scheme to solve convex functional optimization problems
under the conditions in which \gls{IoE} devices communicate with their adjacent devices
and update prediction functions without a coordinator.
\item
\begin{revhl}{1-1}
We analyzed the convergence of the proposed meta-algorithm in a function space without the knowledge of the other devices' \gls{ML} representations.
Spectral graph theory is used to prove convergence with non-\gls{IID} data in a decentralized manner.
We also revealed how network connectivity affects the convergence rate by extending the spectral graph theory to function spaces.
\end{revhl}
\item
\begin{revhl}{1-2}
We develop a \gls{CMFD} for \glspl{NN} as an implementation of the meta-algorithm.
Even though a distributed learning of \glspl{NN} is a non-convex parameter optimization problem,
our algorithm leverages a distillation scheme to update parameters
so that prediction functions trace the trajectory of solving convex function optimization.
\end{revhl}
\item
\Gls{CMFD} works in a fully distributed manner over multi-hop networks with non-\gls{IID} data.
Owing to distillation, \gls{CMFD} can be adopted even when the \gls{NN} architectures among the devices differ,
which is suitable for \gls{IoE} devices with various computational resources.
To the best of our knowledge, conventional \gls{FL} algorithms cannot have all these features simultaneously.
\end{enumerate}

The rest of this paper is organized as follows:
Related works are discussed in Section~\ref{sec:related}.
The problem definition is described in Section~\ref{sec:problem}.
The proposed algorithm is explained in Section~\ref{sec:algorithm}
and is evaluated in Section~\ref{sec:evaluation}.
Finally, conclusions are drawn in Section~\ref{sec:conclusion}.

\section{Related works} \label{sec:related}
\subsubsection*{Decentralized learning algorithms in multi-hop networks}
Multi-hop \gls{FL} for \gls{IoE} devices was introduced in \cite{lalitha2019peer,savazzi2020federated,lian2017can,sato2020network}.
Savazzi \etal \cite{savazzi2020federated} developed a multi-hop \gls{FL} algorithm that shares both model updates and gradients to improve convergence.
Lalitha \etal \cite{lalitha2019peer} proposed a Bayesian approach to estimate the global model and tackle non-\gls{IID} problems
by updating and aggregating beliefs with neighbors.
However, the evaluations were limited to a few nodes, and it is still uncertain whether the algorithm converges with many devices.
Lian \etal \cite{lian2017can} proposed parallel \gls{SGD} in multi-hop topologies
and showed that \gls{DPSGD} accelerates learning when the channel capacity of the centralized server is a bottleneck because of narrow bandwidths.
Sato \etal \cite{sato2020network} extended \gls{DPSGD} to apply it to wireless networks by considering the channel capacity of each link.
These algorithms are operated and analyzed in a parameter space,
fundamentally requiring identical \gls{NN} model architectures among all devices,
which decreases the flexibility of \gls{FL} systems.
In contrast to these algorithms, we utilize distillation to aggregate \gls{NN} models to realize model-free \gls{FL}.

\subsubsection*{Distillation}
While a typical \gls{FL} must send all the parameters of the \gls{NN},
distillation-based \gls{FL} sends only the output of the prediction function, which usually has smaller dimensions than the \gls{NN} parameters.
Therefore, \gls{FD} is proposed to realize communication-efficient algorithms \cite{oh2020mix2fld,ahn2019wireless,itahara2021distillation,chang2019cronus,jeong2018communication}.
Mix2FLD proposed in \cite{oh2020mix2fld} uses \gls{FL} for downlink and \gls{FD} for uplink to address the uplink-downlink capacity asymmetry.
A hybrid \gls{FD} was proposed in \cite{ahn2019wireless} to compensate for the performance gap between \gls{FL} and \gls{FD} by using covariate information among the devices.
In \cite{itahara2021distillation}, unlabeled open data are leveraged to realize semi-supervised \gls{FD}.
Jeong \etal \cite{jeong2018communication} proposed \gls{FAug} by extending \gls{FD}.
\Gls{FAug} leverages a \gls{GAN} to generate \gls{IID} datasets while preserving privacy.
\Gls{FAug} is extended to MultFAug in \cite{jeong2019multi} to be used in multi-hop networks.
MultFAug sends \gls{NN} model information via multi-hop networks, but model aggregation is centralized.
A centralized algorithm cannot be scaled to the number of devices because bottleneck problems arise when aggregating \gls{NN} models.

In contrast to \cite{oh2020mix2fld,ahn2019wireless,itahara2021distillation,chang2019cronus,jeong2018communication}, which use central units,
\cite{anil2018large,zhang2018deep} applied distillation to decentralized learning.
These works show that speed and cost-efficiency are advantages of distillation-based decentralized learning.
However, the evaluations in \cite{anil2018large,zhang2018deep} assume fully connected topologies and \gls{IID} training data,
which are not found in \gls{IoE} systems.
In addition, if each device must share data with all the other devices, the amount of transferred data increases with the number of devices,
consequently, the fully connected algorithm cannot be applied to systems with a large number of devices.
Our algorithm can be applied to devices connected via multi-hop networks, which is useful for \gls{IoE} systems.

Cronus \cite{chang2019cronus}, FedMD \cite{li2019fedmd}, and FedDF \cite{lin2020ensemble}
leveraged distillation to realize heterogeneous \gls{FL} where different \gls{NN} models can be used,
whereas the above-mentioned studies utilized distillation to reduce communication costs and improve performance.
Unlike these algorithms that require a central server,
the proposed algorithm realizes decentralized \gls{FL} with heterogeneous models.

\subsubsection*{Distributed optimization}
Distributed optimization problems have been discussed for \gls{WSN}, and many associated algorithms have been developed.
One of these algorithms is called \gls{CDO},
which can solve convex optimization problems typically designed as follows:
\begin{align}
  \minimize_{\theta\in\Real^n} \sum_i F_i(\theta), \label{eq:distopt}
\end{align}
where the function $F_i$ is only known to agent $i$, and each agent exchanges information with only one-hop neighbors \cite{nedic2009distributed,johansson2010randomized,colin2016gossip}.
With \gls{CDO}, agents individually update parameters $\theta$ by gradient descent and then share the parameters with their neighbors.
These steps are iterated until the parameters converge to the optimal solution.
If optimization problems are convex, convergence to the global optimum is proved in \cite{nedic2009distributed,johansson2010randomized,colin2016gossip}.
\gls{CDO} is also applied to multi-hop \gls{ML} by regarding $F_i$ in (\ref{eq:distopt}) as a loss function with local private data and $\theta$ as the \gls{NN} parameters
\cite{lalitha2019peer,savazzi2020federated,sato2020network}.
\Gls{ADMM} is also applied to decentralized \gls{ML}.
Elgabli \etal proposed a \gls{GADMM} to solve convex \gls{ML} tasks in \cite{elgabli2020gadmm}.
Niwa \etal proposed \gls{PDMM} \cite{niwa2020edge}, which shows a strong convergence even in weakly connected networks, for \gls{FL}.

However, training \gls{NN} tasks are non-convex optimization problems, and, therefore, convergence is not guaranteed.
When multiple devices approach the different local optima by gradient descent,
the direction calculated by parameter aggregation can differ from those of the gradient descent updates,
and performance is degraded.
To overcome this problem,
our concept is to perform \textit{\gls{CDO} in a function space} that optimizes not parameters but prediction functions themselves
by regarding a supervised \gls{ML} task as a functional optimization;
that is, $\theta$ and $F_i$ in (\ref{eq:distopt}) become the prediction function $f$ and loss functional $\mathsf{L}_i$ depending on local data, respectively, as follows:
\begin{align}
  \minimize_{f\in\FnSpc} \sum_i \mathsf{L}_i(f),
\end{align}
where $\mathcal{F}$ denotes the function space.
It is expected that \gls{CDO} in a function space can decentralize the multi-hop \gls{FL} solution
because the commonly used \gls{ML} criteria, \gls{MSE} and \gls{KL} divergence, are convex functionals regarding a prediction function as an argument \cite{van2014renyi},
even though prediction models are non-linear functions with respect to their parameters.
Our algorithm can learn non-\gls{IID} data
because \gls{CDO} is designed to work even when the $\mathsf{L}_i$ among devices differs.

\Gls{CDO}-based algorithms in function spaces were proposed in \cite{koppel2018decentralized,richards2020decentralised,xu2019coke,shen2021distributed}.
These algorithms assume \gls{RKHS}, which is a function space in which any function can be represented by a linear combination of coefficients and kernels.
In \gls{RKHS}, aggregated functions are calculated by adding coefficients and sharing kernels.
Therefore, the convergence of functions can be analyzed in real-vector spaces.
In contrast, we analyze convergence in $L^2$ space without using kernels;
therefore, the analysis is more versatile.
Additionally, we prove the convergence with non-\gls{IID} datasets, which is not considered in \cite{koppel2018decentralized,richards2020decentralised,xu2019coke,shen2021distributed}.
We introduce Radon-Nikodym derivative to represent skewness among local spaces depending on skewed data distributions
to analyze the convergence in non-\gls{IID} situations.
The results indicate how this skewness affects convergence in non-\gls{IID} situations.

\section{Problem statement} \label{sec:problem}
We consider that \gls{IoE} devices with sensors are connected via a multi-hop network,
and each device $i$ can only communicate with its adjacent devices $\Neigh{i}$ in each step.
Device $i$ manages its prediction model $f_i$
and updates $f_i$ with its local dataset $\Dsi$ using a \gls{ML} scheme.
The local dataset should not be shared because it includes private information.
Therefore, instead of sharing dataset $\Dsi$,
the devices share some information representing $f_i$ with their adjacent devices $\Neigh{i}$.
By using the received information, the devices train their prediction model $f_i$
to be optimal even if the local datasets are non-\gls{IID}.
The mathematical problem is formulated as follows:

Let $\InSpc\subseteq\Real^\DimX$ and $\OutSpc\subseteq\Real^\DimY$
be an input space (or feature space) and an output space, respectively.
We denote by $f : \InSpc \to \OutSpc$ as a prediction function,
and $\trueF$ as the ground truth function.
Let $l(\bm{y}, \bm{y}') : \OutSpc\times\OutSpc \to \Real$ be the cost function of each element $\bm{y},\bm{y}'\in\OutSpc$.
When the \gls{MSE} criterion is used, $l(\bm{y},\bm{y}')$ is defined as $l(\bm{y},\bm{y}')\coloneqq \lv \bm{y}-\bm{y}' \rv^2$,
and when the \gls{KL} divergence is used, it is defined as $l(\bm{y},\bm{y}')\coloneqq \sum_{m=1}^{\DimY}y_m\log \frac{y_m}{y'_m}$,
where $y_m$ denotes the $m$th element of vector $\bm{y}$.
Here, a global loss functional $\Lmu(f)$
and local loss functional $\Lmui(f)$ for each user $i\in\Users$ are defined as follows:
\begin{align}
  \Lmu(f) &\coloneqq \int_\InSpc l\(f(\bm{x}),\trueF(\bm{x})\) \dmu, \label{eq:global_loss} \\
  \Lmui(f) &\coloneqq \int_\InSpc l\(f(\bm{x}),\trueF(\bm{x})\) \dmui, \label{eq:local_loss}
\end{align}
where $\mu$ and $\mu_i$ represent a global and local probability measure on $\InSpc$, respectively,
and $\Users$ represents a set of users.
We use an integral form of risk minimization \cite{vapnik1992principles}
because the meta-algorithm is designed to work in a function space that does not depend on a dataset.
We denote by $\LSpc$ the $L^2$ space with respect to $\mu$
and by $\inner{\cdot}{\cdot}{\LSpc}$ an inner product and norm are defined as
$\inner{f}{g}{\LSpc} \coloneqq \int_\InSpc \inner{f(\bm{x})}{g(\bm{x})}{\OutSpc}\,\dmu$ and
$\Dist{f}_{\LSpc} \coloneqq \sqrt{\inner{f}{f}{\LSpc}}$, respectively,
where $\inner{\cdot}{\cdot}{\OutSpc}$ represents the inner product in vector space $\OutSpc$.

This paper focuses on convex loss functionals in the function space $\LSpc$
because the commonly used criteria (e.g., \gls{MSE} and \gls{KL} divergence \cite{van2014renyi}) are convex.
\begin{assumption} \label{asm:convex}
$\Lmu(f)$ and $\Lmui(f)$ satisfy the following inequalities for all $f_1$, $f_2$, and $t\in\[0,1\]$:
\begin{align}
  \Lmu(tf_1 + (1-t)f_2) &\le t \Lmu(f_1) + (1-t) \Lmu(f_2), \\
  \Lmui(tf_1 + (1-t)f_2) &\le t \Lmui(f_1) + (1-t) \Lmui(f_2).
\end{align}
\end{assumption}

To discuss the convergence of the gradient descent algorithm,
both the global and local loss functionals are assumed to be Lipschitz continuous.
In contrast to \cite{lian2017can}, which assumes Lipschitz gradient continuity in a parameter space,
we assume Lipschitz continuity in a function space.
\begin{assumption} \label{asm:lipschitz}
  $\Lmui(f)$ satisfies the Lipschitz condition for all $f_1$ and $f_2$:
  \begin{align}
    &|\Lmui(f_1) - \Lmui(f_2)| \le \Lip{i}\Dist{f_1 - f_2}_{\Lmui},
  \end{align}
  where $\Lip{i}$ denotes the Lipschitz constant of $\Lmui(f)$.
\end{assumption}
This condition is similar to the assumption of the gradient limit considered in \cite{li2020convergence}
because the Lipschitz constant corresponds to the maximum of the gradients.
Note that not all tasks satisfy this assumption;
e.g., $\Lmui$ is defined as the squared $L^2$ norm, and the trained functions are not limited to probability measures.
However, gradient methods can practically be applied even when this assumption is not satisfied \cite{boyd2003subgradient}.

In the considered system,
it is assumed that the ground truth function is common among users,
but the probability measures $\mu_i$ vary across users, that is, users have non-\gls{IID} data.
This setting is referred to as ``feature distribution skew'' and ``label distribution skew'' in \cite{kairouz2019advances}.
Considering the meaning of the global probability measure,
it is empirically represented as 
$\mu = \frac{1}{\NumU} \sum_{i\in\Users} \mu_i$, 
where $\NumU$ represents a number of users.
Using this, global loss functional $\Lmu(f)$ can be written as 
$\Lmu(f) = \frac{1}{\NumU}\sum_{i\in\Users}\Lmui(f)$.
It is also notable that $\mu_i$ is absolutely continuous with respect to $\mu$ ($\mu_i \ll \mu$)
and there exists Radon-Nikodym derivative $\rn_i=\frac{\mathrm{d}\mu_i}{\mathrm{d}\mu}$ for all $i\in\Users$.
We denote by $\maxRn$ the supremum of $\rn_i(\bm{x})$, that is, $\maxRn\coloneqq \sup_{x\in\InSpc} \rn_i(\bm{x})$.
This value is regarded as a metric of non-\gls{IID}ness because it takes the minimum value of $1$ when $\mu_i$ and $\mu$ are the same.

In the \gls{ML} context, because the true loss functional $\Lmu(f)$ cannot be obtained,
the following approximation, called empirical risk, is used:
\begin{align}
  \Lmu(f) &\approx \frac{1}{\Card{\Ds}}\sum_{(\bm{x},\bm{y})\in\Ds} l\(f(\bm{x}), \bm{y}\) \nonumber \\
          &= \frac{1}{\NumU}\sum_{i\in\Devices} \frac{1}{\Card{\Dsi}} \sum_{(\bm{x},\bm{y})\in\Dsi} l\(f(\bm{x}), \bm{y}\),
\end{align}
where $\Dsi$, $\Ds$, and $\Card{\cdot}$ respectively denote a labeled dataset of each device,
the union of all the data sets $\Dsi$,
and the cardinality of a set.
Here, we assume $\Card{\Dsi}=\frac{1}{\NumU}\Card{\Ds}$ for all $i\in\Users$ for simplicity,
but it is easily extended to general cases.
When $f$ is represented by a \gls{NN} model with parameter vector $\bm{w}$,
the optimization problem is defined in a parameter space instead of the function space as follows:
\begin{align}
  \minimize_{\bm{w}\in{\Real^{\DimW}}} \quad \Lmu(f(\cdot;\bm{w})), \label{eq:ml_weight}
\end{align}
where $\DimW$ denotes the number of parameters.
To realize \gls{FL} in multi-hop networks,
this optimization problem should be solved under the constraints that each device $i$ shares information only with its adjacent devices denoted by $\Neigh{i}$.

\Gls{CDO} was applied in \cite{savazzi2020federated,lalitha2019peer} to solve (\ref{eq:ml_weight}) in a distributed manner.
However, such algorithms do not work efficiently
because the objective functions are not convex in a parameter space,
and \gls{CDO} cannot be generally applied to non-convex optimization problems.
Thus, \gls{CDO} in a parameter space can achieve only low performance in the \gls{ML} context,
especially when the local data is non-IID.
In contrast to conventional works that optimize parameters $w$,
we develop an algorithm that optimizes $f$ directly.
Considering that \gls{CDO} achieves an optimal solution when the objective function is convex,
the following optimization problem is expected to be solved,
\begin{align}
  \minimize_{f\in\LSpc} \quad \Lmu(f). \label{eq:ml_function}
\end{align}
The following section explains how to solve this optimization problem
and prove the convergence of the proposed algorithm. 

\begin{figure}[!t]
  \begin{algorithm}[H]
    \caption{Meta-algorithm of consensus-based multi-hop FL in a function space}
    \label{alg:metafl}
    \begin{algorithmic}[1]
      \Require Loss functional: $\Lmui(\cdot)$,


               Probability density function: $\mu_i, \mu$

               Learning and sharing rate: $\lr_t, \lrc$
      \While{not converged}
        \ForAll{device $i=1,\ldots,\NumU$}
          \State $d_i^t \leftarrow \subDifLmui$ \label{algln:meta_local_grad}
          \State $g_i^{t+1} \leftarrow f_i^t - \lr_t d_i^t \rn_i $ \label{algln:meta_grad_descent}
          \State Send $g_i^t$ to neighbor devices \label{algln:meta_share}
        \EndFor
        \ForAll{device $i=1,\ldots,\NumU$}
          \State $f_i^{t+1} \leftarrow g_i^{t+1} - \lrc \NumNeigh_i \(g_i^{t+1} - \frac{1}{\NumNeigh_i}\sum_{j\in\Neigh{i}}g_j^{t+1}\)$ \label{algln:meta_aggregation}
        \EndFor
      \EndWhile
    \end{algorithmic}
  \end{algorithm}
\end{figure}

\section{Consensus-based multi-hop federated learning} \label{sec:algorithm}
\subsection{Meta-algorithm in a function space}
First, we explain a meta-algorithm in a function space $\LSpc$ to solve (\ref{eq:ml_function})
and prove that the prediction functions of all devices converge to the optimal solution.
Subsequently, distillation is introduced in Sec.~\ref{sec:alg_distillation} to apply the proposed meta-algorithm to the \gls{NN} models.

To solve (\ref{eq:ml_weight}),
existing \gls{CDO}-based \gls{FL} schemes in a parameter space \cite{savazzi2020federated,lian2017can,lalitha2019peer,sato2020network}
update local models using the following equations:
\begin{align}
  \hat{\bm{w}}_i^{t+1} &\leftarrow \bm{w}_i^t - \lr_t \nabla_{\bm{w}}\sum_{(x,y)\in\Dsi} l(f(\bm{x};\bm{w}_i^t),\bm{y}), \label{eq:conventional_sgd}\\
  \bm{w}_i^{t+1} &\leftarrow \hat{\bm{w}}_i^{t+1} - \lrc \sum_{j\in\Neigh{i}}\(\hat{\bm{w}}_i^{t+1} - \hat{\bm{w}}_j^{t+1}\), \label{eq:conventional_cdo}
\end{align}
where $\bm{w}_i^t$, $\hat{\bm{w}}_i^t$, and $\lr_t$ respectively represent
the parameters of device $i$ at epoch $t$, the temporal parameters, and the learning rate.
We assume that $\lr_t$ monotonically decreases and, therefore, $\lr_t$ satisfies $\lr_t \le \lr_1$.
The coefficient $\lrc$ of the second term of (\ref{eq:conventional_cdo}) is referred to as the sharing rate, which adjusts the convergence among devices.
To operate (\ref{eq:conventional_sgd}) and (\ref{eq:conventional_cdo}),
$\hat{\bm{w}}_i^t$ should be shared among adjacent devices,
and thus wide communication bandwidths are required
when a \gls{NN} model has many layers.
In addition, the \gls{CDO} is not guaranteed to reach the optimal solution for non-convex optimizations.

To tackle these problems, we replace parameter optimization by function optimization as follows:
\begin{align}
  g_i^{t+1} &\leftarrow f_i^t - \lr_t d_i^t \rn_i, \label{eq:meta_grad} \\
  f_i^{t+1} &= g_i^{t+1} - \lrc \NumNeigh_i \(g_i^{t+1} - \frac{1}{\NumNeigh_i}\sum_{j\in\Neigh{i}}g_j^{t+1}\), \label{eq:meta_consensus}
\end{align}
where $\NumNeigh_i$ represents the number of neighbors of user $i$, defined by $\NumNeigh_i\coloneqq\Card{\Neigh{i}}$.
Here, $\grad{i}{t}\in\subDifLmui$ represents a \gls{Frechet} subgradient of the local loss functional $\Lmui(f)$,
which satisfies the following inequality \cite{ambrosio2008gradient}:
\begin{align}
  \forall h\in\LSpc[\mu_i], \inner{\grad{i}{t}}{h - f_i^t}{\LSpc[\mu_i]} \leq \Lmui(h) - \Lmui(f_i^t). \label{eq:subdifferential}
\end{align}
In (\ref{eq:meta_grad}), $\rn_i=\frac{\mathrm{d}\mu_i}{\mathrm{d}\mu}$
represents the skewness of subgradient $\grad{i}{t}$, which is caused by mini-batch algorithms executed in $\LSpc[\mu_i]$ not in the global space $\LSpc$.
We assume that $\lrc$ in (\ref{eq:meta_consensus}) satisfies $0<\lrc\le\frac{1}{2\MaxD}$ where $\MaxD$ denotes the maximum degree of the network graph.
This condition is required to guarantee convergence, as explained in Sec.~\ref{sec:analysis}.

Algorithm~\ref{alg:metafl} shows pseudocode that applies the above operation.
In each epoch, the devices update their local prediction functions $f_i^t$ by (\ref{eq:meta_grad}) (steps~\ref{algln:meta_local_grad}--\ref{algln:meta_grad_descent}).
After updating them, the devices share temporal functions $g_i^{t+1}$ with the adjacent devices $\Neigh{i}$ (step~\ref{algln:meta_share}).
Then, the devices aggregate the received models and calculate the weighted average of the prediction functions (step~\ref{algln:meta_aggregation}).
After iterating these steps, all the local prediction functions converge to the optimal solution $f^{\star}$ of (\ref{eq:ml_function}).

\subsection{Convergence analysis of the meta-algorithm} \label{sec:analysis}
The convergence analysis is divided into two parts.
First, an upper bound is obtained of the distance between the local prediction functions $f_i^t$ and their average $\MeanF_t$.
Then, we discuss the limits of $\Lmu(\MeanF_t)$ as $t\to\infty$.

In order to mathematically analyze the convergence of the meta-algorithm, we define a federated function $\bm{f}_t$ as a tuple of prediction functions:
\begin{align}
  \bm{f}_t &\coloneqq \(f_{1}^{t}, \ldots, f_{\NumU}^{t}\)^\Trans \in \FedSpc \coloneqq\(\LSpc\)^\NumU,
\end{align}
where $\[\cdot\]^\Trans$ denotes the transpose of $\[\cdot\]$.
$\FedSpc$ is the $\NumU$-ary Cartesian power of $\LSpc$
with the inner product $\inner{\bm{a}}{\bm{b}}{\FedSpc} \coloneqq \sum_{i=1}^{\NumU} \inner{a_i}{b_i}{\LSpc}$
and norm $\lv \bm{a} \rv_{\FedSpc} \coloneqq \sqrt{\inner{\bm{a}}{\bm{a}}{\FedSpc}}$
where $a_i$ and $b_i$ are the $i$th elements of the federated functions $\bm{a}$ and $\bm{b}\in\FedSpc$, respectively.
The addition and scalar multiplication of federated functions are calculated in an element-wise manner.
The product of the  matrices and federated functions is defined similarly to that of matrices and numerical vectors,
that is, the $i$th element of $A\bm{a}\in\FedSpc$ is defined as $\sum_{j=1}^{\NumU}A_{ij}a_j$,
where $A_{ij}$ denotes the $(i,j)$ element of matrix $A$.

The following lemma expresses the relation between the induced norm of real matrices and the federated functions.
\begin{lemma} \label{lemm:inducednorm}
  Let $\Dist{A}$ be the induced norm of matrix $A (\in\Real^{\NumU\times \NumU})$
  corresponding to the 2-norm of real vectors,
  i.e.,
  \begin{align}
    \Dist{A}\coloneqq\max \left\{\Dist{A\bm{v}}_{\Real^\NumU}/\Dist{\bm{v}}_{\Real^\NumU} \mid \bm{v}\in\Real^\NumU, \bm{v} \neq\bm{0} \right\}. \label{eq:def_inducednorm}
  \end{align}
  Then, $\Dist{A}$ is also the induced 2-norm of the federated functions, that is, $\Dist{A}$ satisfies
  \begin{align}
    \Dist{A} = \max \left\{\Dist{A\bm{a}}_{\FedSpc}/\Dist{\bm{a}}_{\FedSpc} \mid \bm{a}\in\FedSpc, \bm{a} \neq\bm{0} \right\}.
  \end{align}
\end{lemma}
\begin{IEEEproof}
  See Appendix \ref{sec:app_inducednorm}.
\end{IEEEproof}
This lemma enables the convergence of Algorithm~\ref{alg:metafl} in $\FedSpc$ to be analyzed similarly to that in $\Real^\NumU$.

Using the federated functions, (\ref{eq:meta_grad}) and (\ref{eq:meta_consensus}) can be rewritten as follows:
\begin{align}
  \bm{g}_{t+1} &\leftarrow \bm{f}_{t} - \lr_t \bm{d}_t, \label{eq:meta_grad_all} \\
  \bm{f}_{t+1} &\leftarrow \bm{g}_{t+1} - \lrc \Laplacian \bm{g}_{t+1} = (I-\lrc \Laplacian) \bm{g}_{t+1}, \label{eq:meta_consensus_all}
\end{align}
where $\bm{g}_t$, $\bm{d}_t$, $I$, and $\Laplacian$ denote federated functions the $i$th elements of which are $g_i^t$ and $d_i^t \rn_i$,
the identity matrix, and the Laplacian matrix of the network graph, respectively.
The $(i,j)$th element of $\Laplacian$ is $\Card{\Neigh{i}}$ if $i=j$, -1 if $j\in\Neigh{i}$, or 0.
Let $\eigen_2$ denote the second minimum eigenvalue of $\Laplacian$, which is also known as algebraic connectivity \cite{fiedler1973algebraic}.
Generally, consensus algorithms converge faster if the value increases.

We denote by $\bar{\bm{a}}$ a mean federated function the elements of which are the mean of functions $a_i$,
which is calculated as $\bar{\bm{a}} \coloneqq \frac{1}{\NumU}\Ones \bm{a}$,
where $\Ones$ represents an $\NumU\times \NumU$ matrix of ones.
Using federated functions, the root-mean-square distance between the local prediction functions and their mean is expressed as follows:
\begin{align}
  \MeanDist_t \coloneqq \sqrt{\frac{1}{n}\sum_{i=1}^{n} \Dist{f_i^t - \MeanF_t}_{\LSpc}^2} = \frac{1}{\sqrt{n}} \Dist{\bm{f}_t - \FedMeanF_t}_{\FedSpc},
\end{align}
where $\MeanF_t$ denotes the mean of the $f_i^t$.
Based on Lemma~\ref{lemm:inducednorm},
the upper bound of $\MeanDist_t$ can be similarly obtained using spectral graph theory as if the federated function $\bm{f}_t$ were a numerical vector.
\begin{theorem}\label{th:convdist}
  If the network graph is connected, the sharing rate satisfies $0 < \lrc \le \frac{1}{2\MaxD}$,
  and Assumptions~\ref{asm:convex} and \ref{asm:lipschitz} hold,
  $\MeanDist_t$ is upper bounded as follows:
  \begin{align}
    \MeanDist_t \le \frac{1}{\sqrt{n}} \initF \convRate^{t-1} + \maxLip\sum_{\tau=1}^{t-1} \lr_{\tau} \convRate^{t-\tau}, \label{eq:th_distbound}
  \end{align}
where we define $\convRate\coloneqq 1-\lrc\eigen_2$ and $\maxLip\coloneqq \max_{i} \left\{\sqrt{\maxRn} \Lip{i}\right\}$.
\end{theorem}
\begin{IEEEproof}
  See Appendix \ref{sec:app_convdist}.
\end{IEEEproof}
Because $\eigen_2$ does not exceed twice the maximum degree $\MaxD$ of the network graph,
if the sharing rate $\lrc$ is selected to satisfy $0<\lrc\le\frac{1}{2\MaxD}$, we have $0\le \convRate < 1$.
Because we assume that the learning rate $\lr_t$ monotonically decreases,
the \gls{RHS} of (\ref{eq:th_distbound}) is bounded as follows:
\begin{align}
  \distAvg_t &\coloneqq \frac{1}{\sqrt{n}} \initF \convRate^{t-1} + \maxLip\sum_{\tau=1}^{t-1} \lr_{\tau} \convRate^{t-\tau} \nonumber \\
             &\le \frac{1}{\sqrt{\NumU}} \initF \convRate^{t-1} + \maxLip \lr_1 \frac{\convRate(1-\convRate^{t-1})}{1-\convRate}.
\end{align}
The limit of $\MeanDist_t$ is bounded by the constant value as follows:
\begin{align}
  \lim_{t\to\infty}\MeanDist_t \le \frac{\lr_1 \convRate \maxLip}{1-\convRate} = \frac{\lr_1 (1-\lrc\eigen_2) \maxLip}{\lrc\eigen_2}. \label{eq:limDist}
\end{align}

Next, we discuss the optimality of $\MeanF_t$.
Because the subgradient method is not a descent method, we consider the best solution yet found in a similar manner as discussed in \cite{boyd2003subgradient}.
The best solution $\bestF$ is defined as
$\bestF \coloneqq \argmin_{\tau=1,\dots,t}\left\{\Lmu\(\MeanF_{\tau}\)\right\}$.

\begin{theorem} \label{th:convopt}
  The best solution $\bestF$ found in $t$-time iterations by Algorithm~\ref{alg:metafl} satisfies the following inequality:
  \begin{align}
    &\Lmu(\bestF) - \Lmu(\trueF) 
           \le \frac{1}{2 \sum_{\tau=2}^{t} \lr_{\tau}}
              \[ \initDist + \maxLip \sum_{\tau=2}^{t}\lr_{\tau}^2 \right. \nonumber \\
    &\qquad    \left. + \Coeff \(1-\convRate^{t-1}\) \(\lr_1 \initF + \sqrt{\NumU} \maxLip \sum_{\tau=1}^{t-1}\lr_{\tau}^2\)\], \label{eq:th_optimality}
\end{align}
where $\initDist$ and $\Coeff$ are defined as $\initDist\coloneqq\Dist{\MeanF_{2} - \trueF}_{\LSpc}^2$
and $\Coeff\coloneqq \frac{4 \maxLip \convRate}{1-\convRate}$, respectively.
\end{theorem}
\begin{IEEEproof}
  See Appendix \ref{sec:app_convopt}.
\end{IEEEproof}
Consider that $\maxLip=\max_{i} \left\{\sqrt{\maxRn} \Lip{i}\right\}$,
and $\maxRn$ is a metric of non-\gls{IID}ness,
which takes the minimum value of $1$ when all the local data distribution $\mu_i$ are the same.
Therefore, the effect of data-distribution skewness on convergence rate is observed in (\ref{eq:th_optimality}).
When the learning rate $\lr_t$ satisfies $\sum_{t=1}^{\infty} \lr_t^2 <\infty$ and $\sum_{t=1}^{\infty} \lr_t = \infty$,
the right-hand side of (\ref{eq:th_optimality}) converges to zero as $t\to\infty$.
In contrast, when a constant learning rate is adopted, i.e., $\lr_t =\lr$ for all $t$,
\begin{align}
  &\lim_{t\to\infty} \Lmu(\bestF) - \Lmu(\trueF) \nonumber \\
  &\quad \le \frac{\lr\maxLip}{2}\(1 + \frac{4 \sqrt{n} \maxLip (1-\lrc\eigen_2)}{\lrc\eigen_2}\). \label{eq:limBest}
\end{align}
Although it is not guaranteed that the sub-gradient method strictly decreases $\Lmu(\MeanF_t)$,
a decreasing series can be obtained by tuning the learning rate.

\subsection{\gls{CMFD} as an implementation for neural networks} \label{sec:alg_distillation}
\begin{figure}[!t]
  \begin{algorithm}[H]
    \caption{Pseudo code of \gls{CMFD}}
    \label{alg:nnfl}
    \begin{algorithmic}[1]
      \Require Prediction function with \gls{NN} parameters $\bm{w}$: $f(\cdot;\bm{w})$,


               Cost function: $l(y,y')$

               Local and shared train data: $\Dsi$, $\sharedD$,

               Learning and sharing rate: $\lr_t, \lrc$
      \While{not converged}
        \ForAll{device $i=1,\ldots,\NumU$}
          \For {minibatch $\Dsm$ in $\Dsi$} \label{algln:local_sgd_bgn}
            \State $\ws^{t+1} \leftarrow \bm{w}_i^t - \lr_t \nabla_{\bm{w}} \displaystyle{\sum_{(\bm{x},\bm{y})\in\Dsm}}l(f(\bm{x};\bm{w}_i^t),\bm{y}) $ \label{algln:local_sgd_grad}
          \EndFor \label{algln:local_sgd_end}
          \ForAll {$x\in\sharedD$} \label{algln:pub_calc_bgn}
            \State $\predy{i,\bm{x}}^{t+1} \leftarrow f(\bm{x};\ws^{t+1})$
          \EndFor
          \State send $\predy{i,\bm{x}}^{t+1}$ to neighbor devices \label{algln:pub_calc_end}
        \EndFor
        \ForAll{device $i=1,\ldots,\NumU$}
          \State receive $\predy{j,\bm{x}}^{t+1}$ from neighbor devices
          \For {minibatch $\Dsm$ in $\sharedD$}
            \State $\bm{w}_i^{t+1} \leftarrow \ws^{t+1} - \lrc \NumNeigh_i \nabla_{\bm{w}} c(\ws^{t+1})$ \label{algln:distil_end} 
          \EndFor
        \EndFor
      \EndWhile
    \end{algorithmic}
  \end{algorithm}
\end{figure}

Because the meta-algorithm works in a function space, practical implementation is required.
We developed a \gls{CMFD} for \glspl{NN} by extending \gls{FL}.
The meta-algorithm directly updates functions using (\ref{eq:meta_grad}) and (\ref{eq:meta_consensus}).
\Gls{CMFD} substitutes these two operations by updating the parameters of the \gls{NN} models, that is, \gls{SGD} and distillation. 
The pseudocode of the proposed algorithm is presented in Algorithm~\ref{alg:nnfl}.

First, we simply substitute the \gls{Frechet} subgradient in (\ref{eq:meta_grad})
with a stochastic gradient in a parameter space (steps~\ref{algln:local_sgd_bgn}--\ref{algln:local_sgd_end}).
Device $i$ updates its local prediction models $f(\cdot;\bm{w}_i)$ using the local dataset $\Dsi$.
Note that the gradient calculated at each local device in step \ref{algln:local_sgd_grad}
is expected to be an approximation of $\grad{i}{t}\rn_i$ rather than $\grad{i}{t}$ in (\ref{eq:meta_grad}),
because $\Dsm$ is a mini-batch of $\Dsi$, which is also exposed to the distribution skew.

\begin{revhl}{1-2}
\Gls{CMFD} utilizes distillation, where devices share the input and output pairs of their prediction models instead of the parameters
to aggregate the prediction models.
This is because parameter averaging (\ref{eq:conventional_cdo}) cannot realize function aggregation (\ref{eq:meta_consensus}) due to the non-convexity of local functions.
\end{revhl}

\begin{figure}[!t]
\centering
\includegraphics[width=\figsizeDistillation]{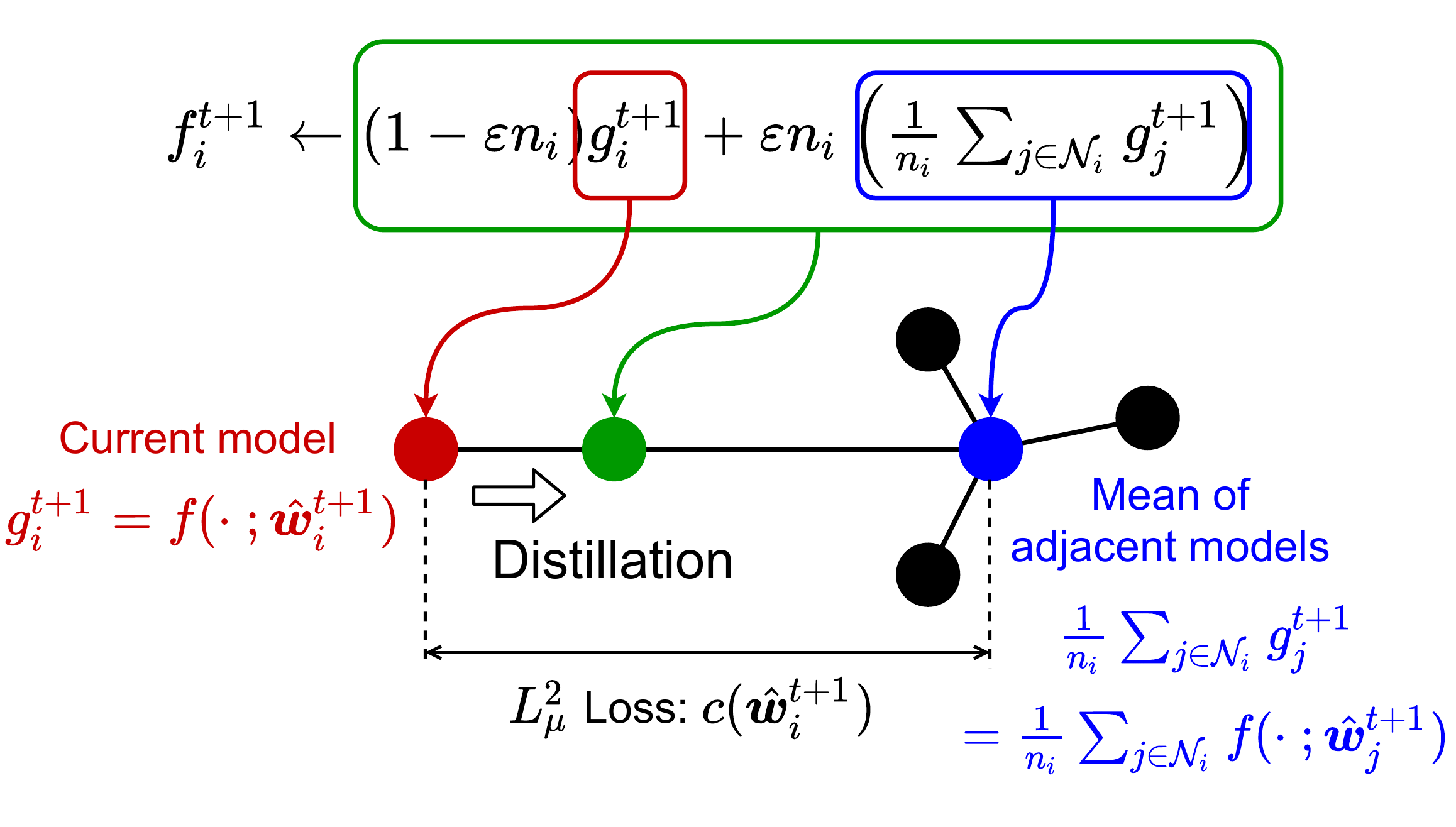}
\caption{
\revhlm{(1-2) }{
Concept of distillation-based function aggregation.
Points in the figure represent the positions of the functions in a function space.
The red point is the position after the local gradient step.
The blue point represents the mean of the functions of the adjacent devices,
which is regarded as a target at distillation steps.
By updating $\ws^{t+1}$ to reduce $c(\ws^{t+1})$,
prediction function approaches the green point.
}
}
\label{fig:distillation}
\end{figure}

\revhlnon{
Fig.~\ref{fig:distillation} shows the concept of distillation-based function aggregation.
Points in Fig.~\ref{fig:distillation} represent the positions of the functions in a function space.
The red, blue, and green points represent
the local prediction function before the aggregation step,
the mean of the prediction functions of the adjacent devices,
and the weighted mean of these two functions, respectively.
Regarding the blue points as a distillation target,
the local prediction function approaches the green points by a gradient descent scheme.
Here, the loss function for distillation $c(\ws^{t+1})$ is defined as follows:
\begin{align}
  c(\ws^{t+1}) &\coloneqq \sum_{(\bm{x},\bm{y})\in\sharedD}\Big\lVert f(\bm{x};\ws^{t+1}) - \frac{1}{\NumNeigh_i}\sum_{j\in\Neigh{i}}\predy{j,\bm{x}}^{t+1}\Big\rVert^2. \label{eq:distillation_loss}
\end{align}
\Gls{MSE} is used as a loss function because it is an empirical representation of the $\LSpc$ distance between two functions.
Fortunately, $c(\ws^{t+1})$ can be calculated without the information of the adjacent devices' models but only with their outputs,
and thus, only the outputs should be shared among the devices.
}

\revhlnon{
The devices calculate outputs $\predy{i,x}^{t+1}$ of public dataset $\sharedD$, and send them to their neighbors $\Neigh{i}$ (steps~\ref{algln:pub_calc_bgn}--\ref{algln:pub_calc_end}).
By using the received values, the devices update their models by distillation
assuming the aggregate outputs $\frac{1}{\NumNeigh_i}\sum_{j\in\Neigh{i}}\predy{j,\bm{x}}^{t+1}$ as new labels
as follows (step~\ref{algln:distil_end}):
\begin{align}
  \bm{w}_i^{t+1} &\leftarrow \ws^{t+1} - \lrc \NumNeigh_i \nabla_{\bm{w}} c(\ws^{t+1}). \label{eq:distillation}
\end{align}
Note that complete distillation at each iteration is not required,
because distillation steps are derived from (\ref{eq:meta_consensus}),
the \gls{RHS} of which is defined as the weighted mean of the local model $g_i^{t+1}$ and $(1/\NumNeigh)\sum_{j\in\Neigh{i}}g_j^{t+1}$.
Therefore, a small update by distillation is sufficient for approximating (\ref{eq:meta_consensus}),
and we performed one epoch for distillation at each iteration in the evaluations in Sec.~\ref{sec:evaluation}.
After sufficient iterations of the \gls{SGD} with local datasets and distillation with public datasets,
all the local models are expected to approach the optimal solution of (\ref{eq:ml_function}),
even though the parameters $\bm{w}_i$ are not forced to be the same values by Algorithm~\ref{alg:nnfl}.
}

We use an additional public dataset, $\sharedD$, as the input data for distillation.
Dataset $\sharedD$ is selected randomly from open data by specifying the category of the input domain,
e.g., animal images for animal recognition tasks and human images for activity recognition tasks.
Although the local datasets $\Dsi$ are non-\gls{IID} because of users' interests,
the public dataset $\sharedD$ is expected to be the empirical representation of the global measure $\mu$.
The selected dataset $\sharedD$ is shared among the devices before the learning algorithm is initiated.
Note that the labels of $\sharedD$ are not required, and thus, they are easy to collect.
If the local dataset $\Dsi$ was used instead of $\sharedD$ in the distillation steps,
\gls{MSE} would be an approximation of $\LSpc[\mu_i]$ rather than $\LSpc$,
and the models would converge to the local optimum in $\LSpc[\mu_i]$.
In addition, local datasets cannot be shared because they may include private information.
Therefore, the public dataset $\sharedD$ is required to aggregate local models in $\LSpc$ by \gls{MSE},
and aggregation in $\LSpc$ realizes the convergence of the models to the optimum in $\LSpc$. 

As mentioned above, \gls{CMFD} is an approximation of the meta-algorithm.
This fact provides us with an intuitive understanding of the convergence property of the \gls{CMFD},
although the convergence is not mathematically guaranteed.
First, the local models as functions approach each other regardless of the values of their parameters.
Second, the convergence of \gls{CMFD} improves if (\ref{eq:meta_grad}) and (\ref{eq:meta_consensus}) are well approximated.
For instance, using a large public dataset should improve performance
because the probability measure of $\sharedD$ approaches $\mu$,
and the distillation becomes a good approximation of (\ref{eq:meta_consensus}).
Third, several characteristics replicate those of \gls{CDO}, such that a strong connection improves convergence.
We also evaluate the convergence property of \gls{CMFD} by simulation in the following section
and confirm that \gls{CMFD} can be applied to various \gls{ML} tasks.

\section{Performance evaluation} \label{sec:evaluation}

\begin{table}[!t]
\caption{Simulation parameters}
\label{tbl:sim_param1}
\centering
\begin{tabular}{cc}
\toprule
Parameters & Values \\
\midrule
Num. devices & 10 \\
Network topology & Ring lattice, BA network \\
\multirow{2}{*}{Num. local data $\Card{\Dsi}$} & 1000 (MNIST, F-MNIST) \\
                                               & 4000 (CIFAR10, CIFAR100) \\
\multirow{2}{*}{Num. public data $\Card{\sharedD}$} & 1000 (MNIST, F-MNIST) \\
                                                    & 10000 (CIFAR10, CIFAR100) \\
Num. labels of local data & 2 \\
Learning rate $\lr$ & 0.01, 0.1 (constant) \\
Sharing rate $\lrc$ & 0.001--6 (constant) \\
Optimizer & SGD \\
Minibatch size & 100 \\
\multirow{2}{*}{Dropout rate (CNN)} & 0.5 (MNIST, F-MNIST) \\
                                    & 0.2 (CIFAR10, CIFAR100) \\
Dropout rate (FC) & 0.1 \\
\bottomrule
\end{tabular}
\end{table}

\begin{figure}[!t]
  \centering
  \subfloat[R1]  {\includegraphics[width=\figsizeTopology]{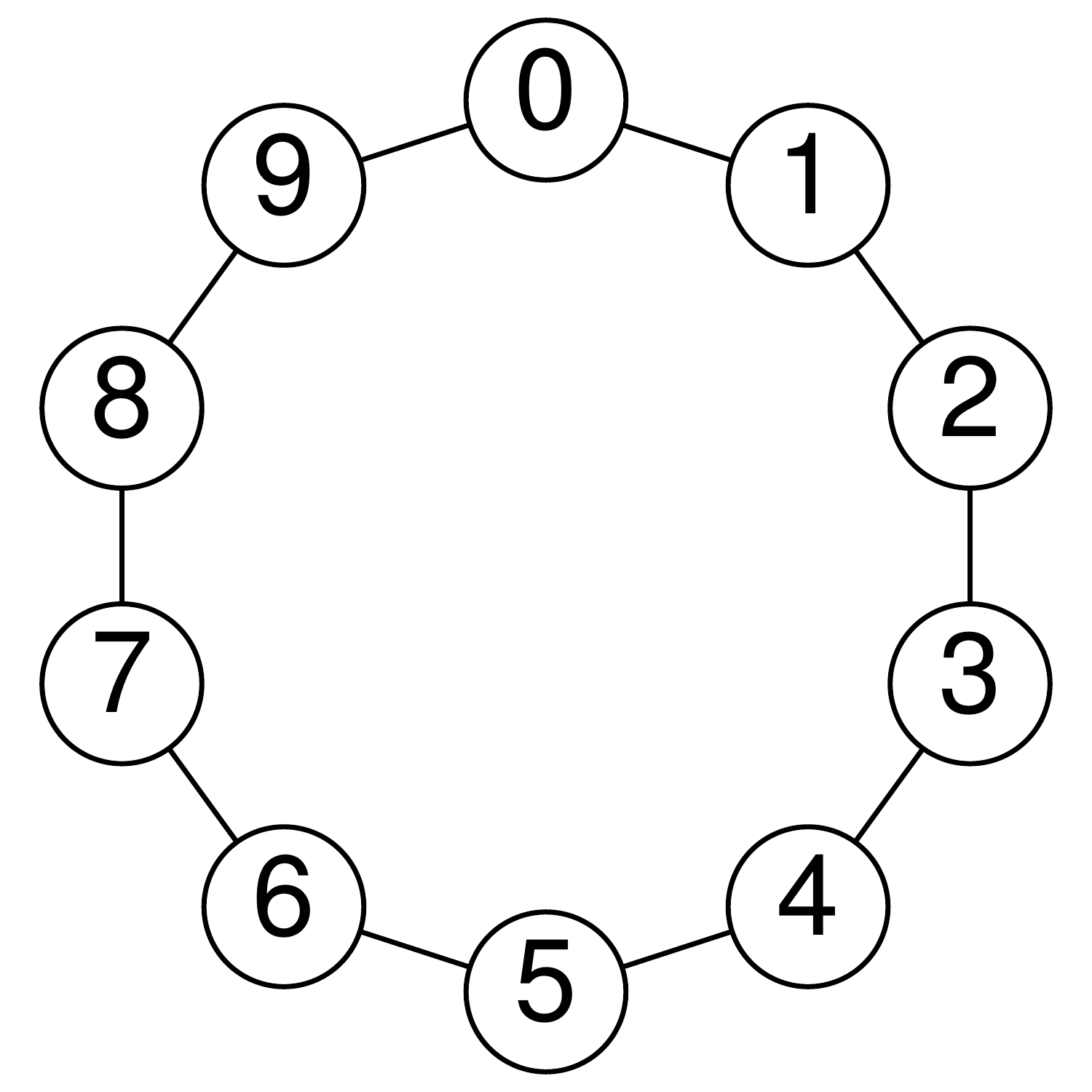}\label{fig:topo_r1}}
  \subfloat[R2]  {\includegraphics[width=\figsizeTopology]{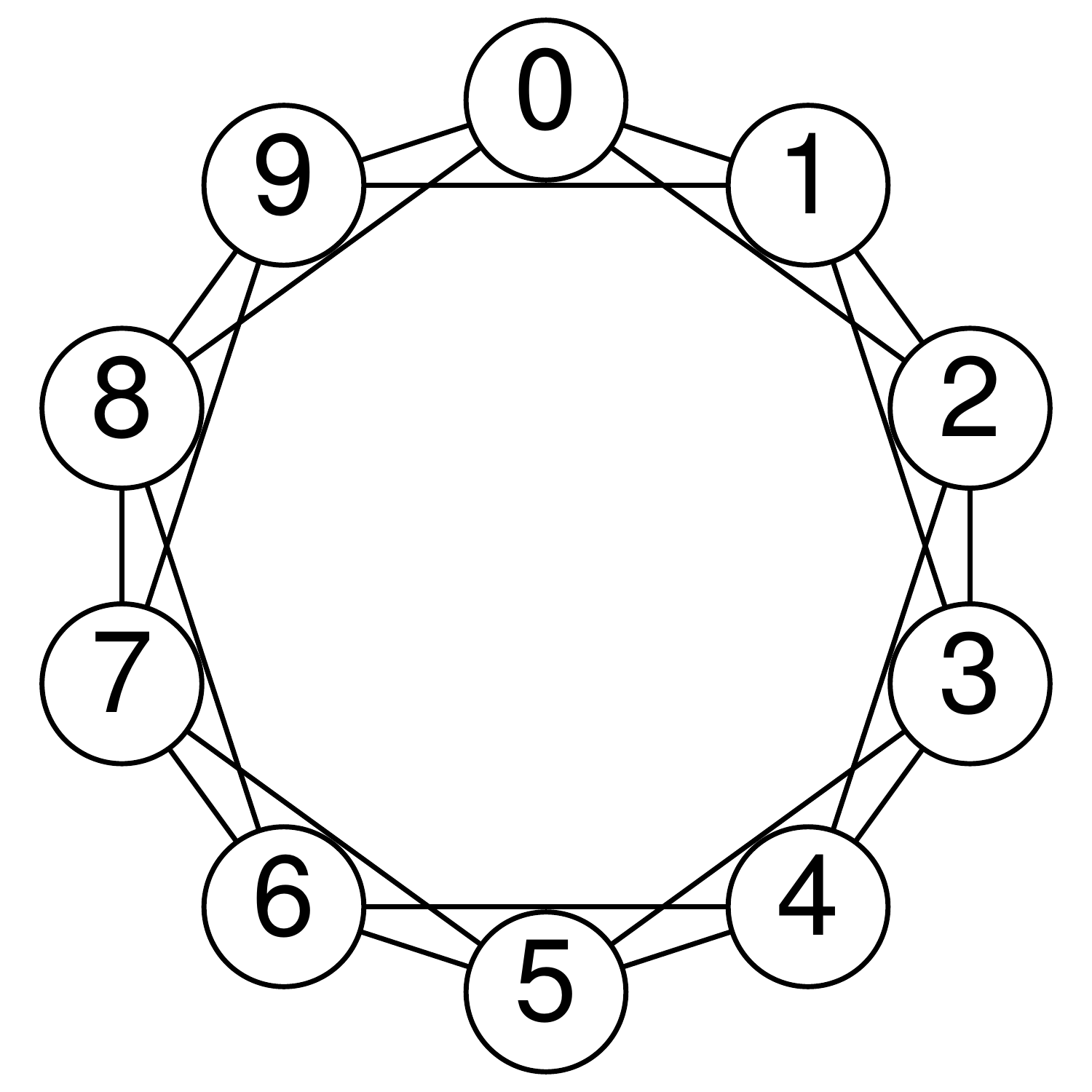}\label{fig:topo_r2}}
  \subfloat[R3]  {\includegraphics[width=\figsizeTopology]{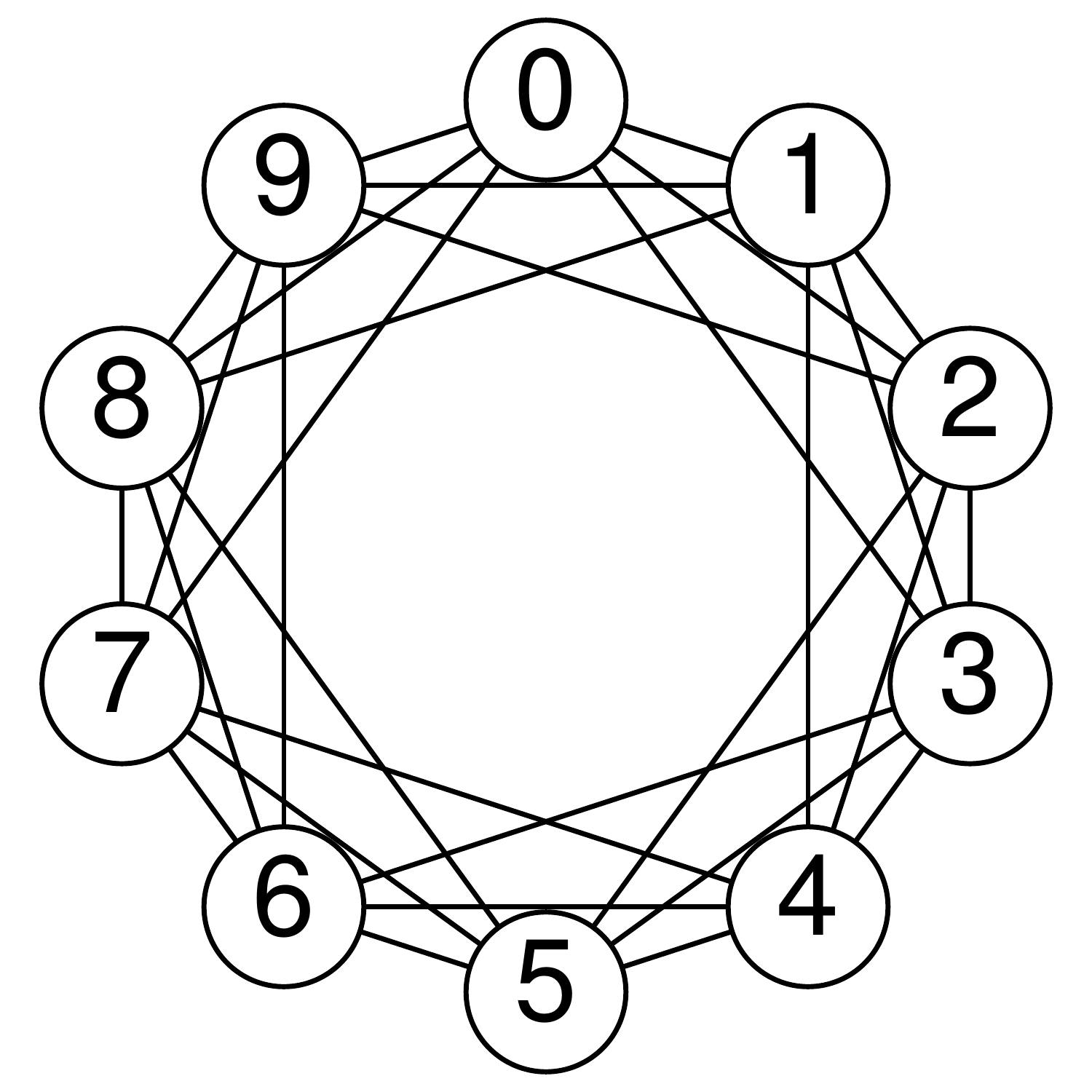}\label{fig:topo_r3}}
  \subfloat[BA1] {\includegraphics[width=\figsizeTopology]{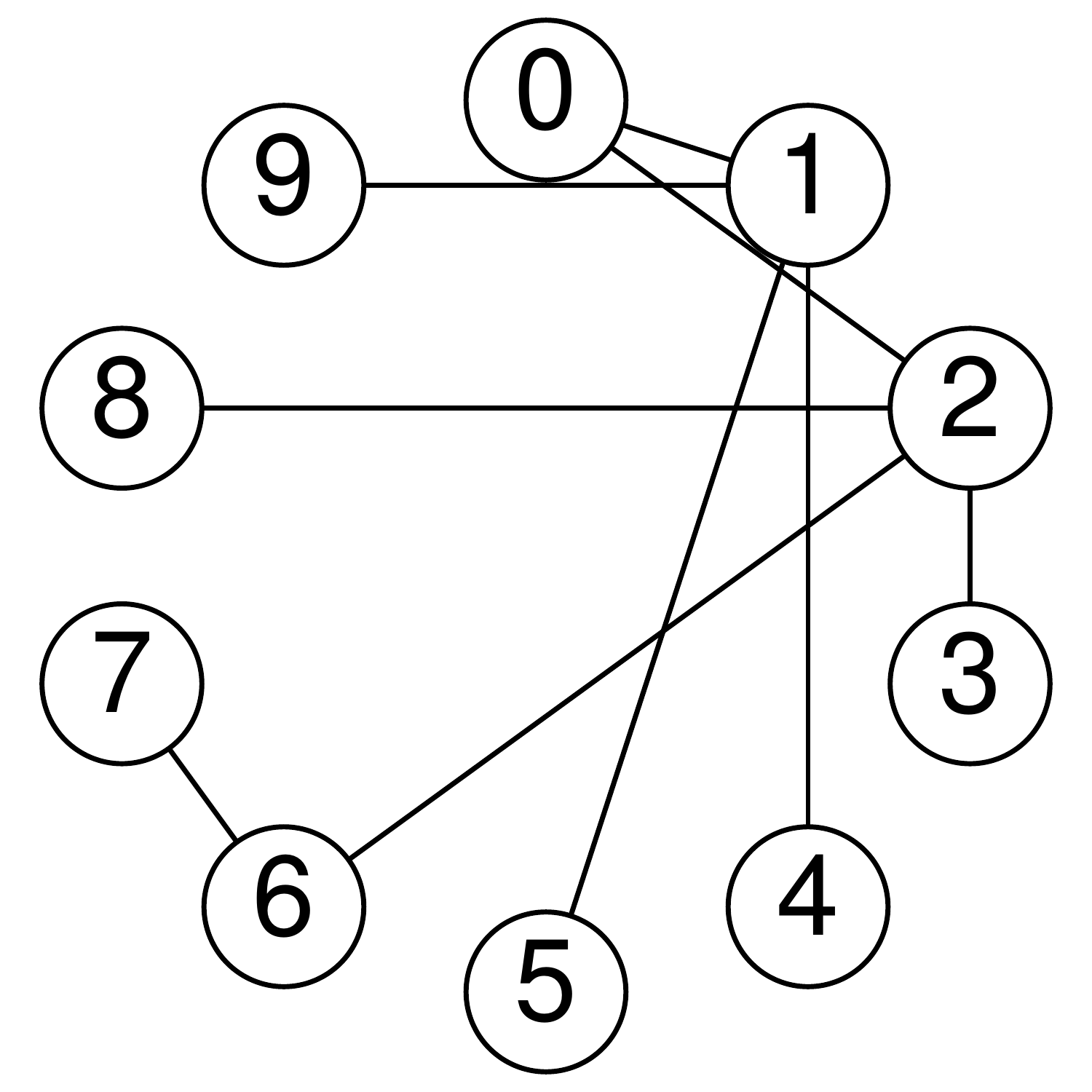}\label{fig:topo_ba1}}
  \subfloat[BA2] {\includegraphics[width=\figsizeTopology]{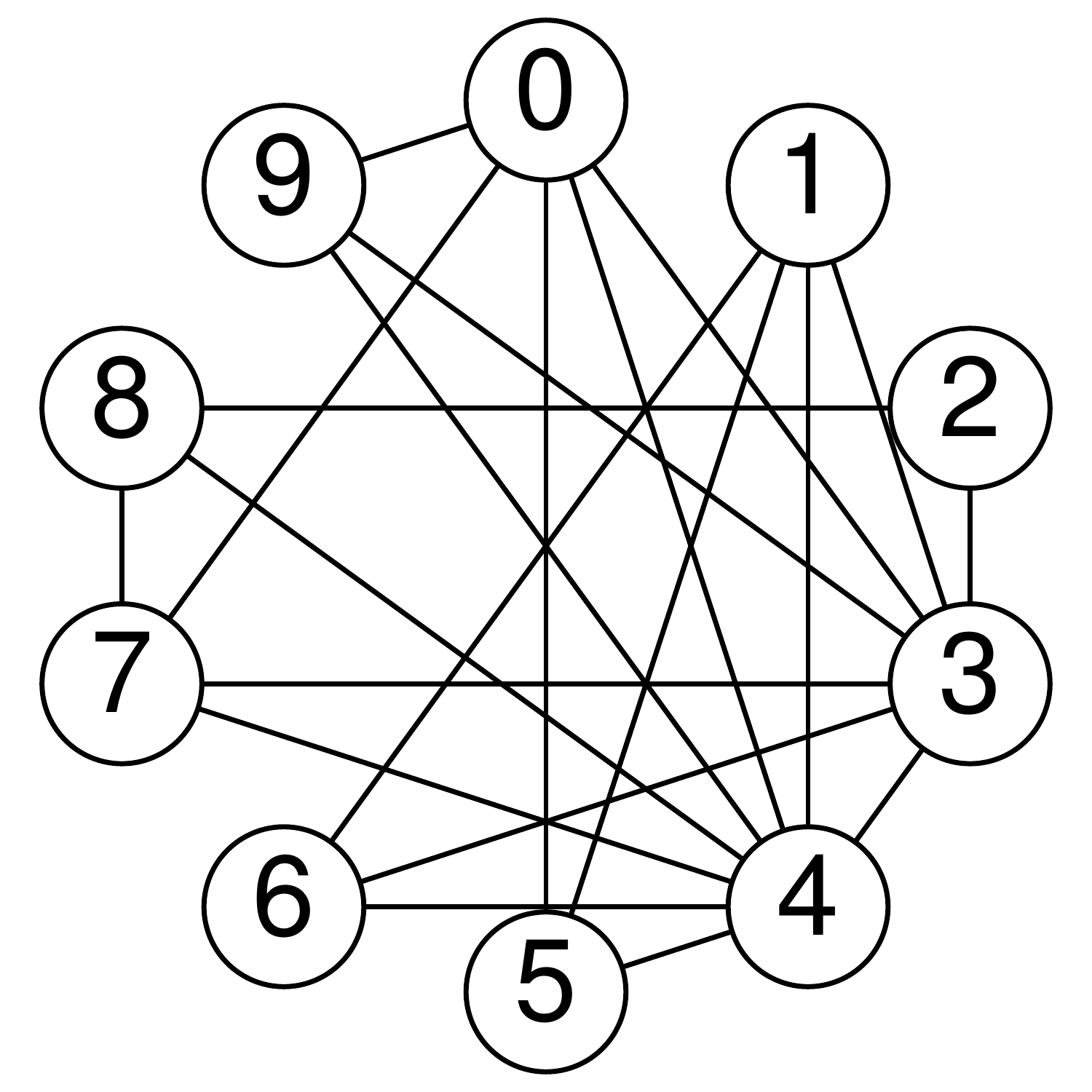}\label{fig:topo_ba2}}
  \caption{Evaluated network topologies. R1, R2, and R3 are ring lattice networks, and BA1 and BA2 are scale-free networks generated by BA algorithm.}
  \label{fig:topologies}
\end{figure}

We evaluated \gls{CMFD} using the public datasets, MNIST \cite{lecun1998gradient}, \gls{FMNIST} \cite{xiao2017fashion}, CIFAR10, and CIFAR100 \cite{krizhevsky2009learning}.
The simulation parameters are presented in Table~\ref{tbl:sim_param1}.
Ten devices were assumed to be connected via multi-hop networks, as shown in Fig.~\ref{fig:topologies}.
We evaluated the R1, R2, and R3 ring lattice networks and BA1 and BA3 scale-free networks
that are occasionally used in \glspl{WSN} for energy efficiency \cite{zhu2009complex}.
The evaluated scale-free networks were generated using the \gls{BA} algorithm \cite{barabasi1999emergence}.
In our simulations, the devices had two types of labels to evaluate the non-\gls{IID} situations.
When evaluating ring networks,
device $\#i$ possessed training data labeled $i$ and $((i+1) \bmod 10)$ for 10-class classification tasks
and $10 i$ through $((10 i + 19) \bmod 100)$ for CIFAR100
assuming a situation where nearby devices had similar data distributions,
where $\bmod$ denotes the modulo operator.
Conversely, when evaluating scale-free networks,
two types of labels were chosen randomly.
One thousand public data samples without ground truth labels were used for the distillation.
The local and public datasets were allowed to duplicate.
The \gls{NN} layer architecture was CNN($32, 5$)-CNN($64, 5$)-FC($512$)-FC($10$)\footnote{CNN($f, k$) and FC($n$) represent the convolution layer with $f$-filters with a kernel size of $k\times k$ and the fully connected layer with $n$-hidden nodes.}.
ReLU was used as the activation function.

We evaluated the performance of the \gls{CDO} in a parameter space
as a parameter-averaging algorithm in multi-hop networks.
With this algorithm, each device updates its local models by \gls{SGD}
similar to \gls{CMFD} (steps~\ref{algln:local_sgd_bgn}--\ref{algln:local_sgd_end} in Algorithm~\ref{alg:nnfl})
and send $\ws^t$ to the adjacent devices.
Then, the device updates its parameters using (\ref{eq:conventional_sgd}) and (\ref{eq:conventional_cdo}).
When evaluating the parameter-averaging method, the initial parameters are set to be equal
to avoid performance deterioration \cite{mcmahan2016communication}.

\subsection{Convergence performance}
\begin{figure}[!t]
\centering
\includegraphics[width=\figsizeConvergenceAll]{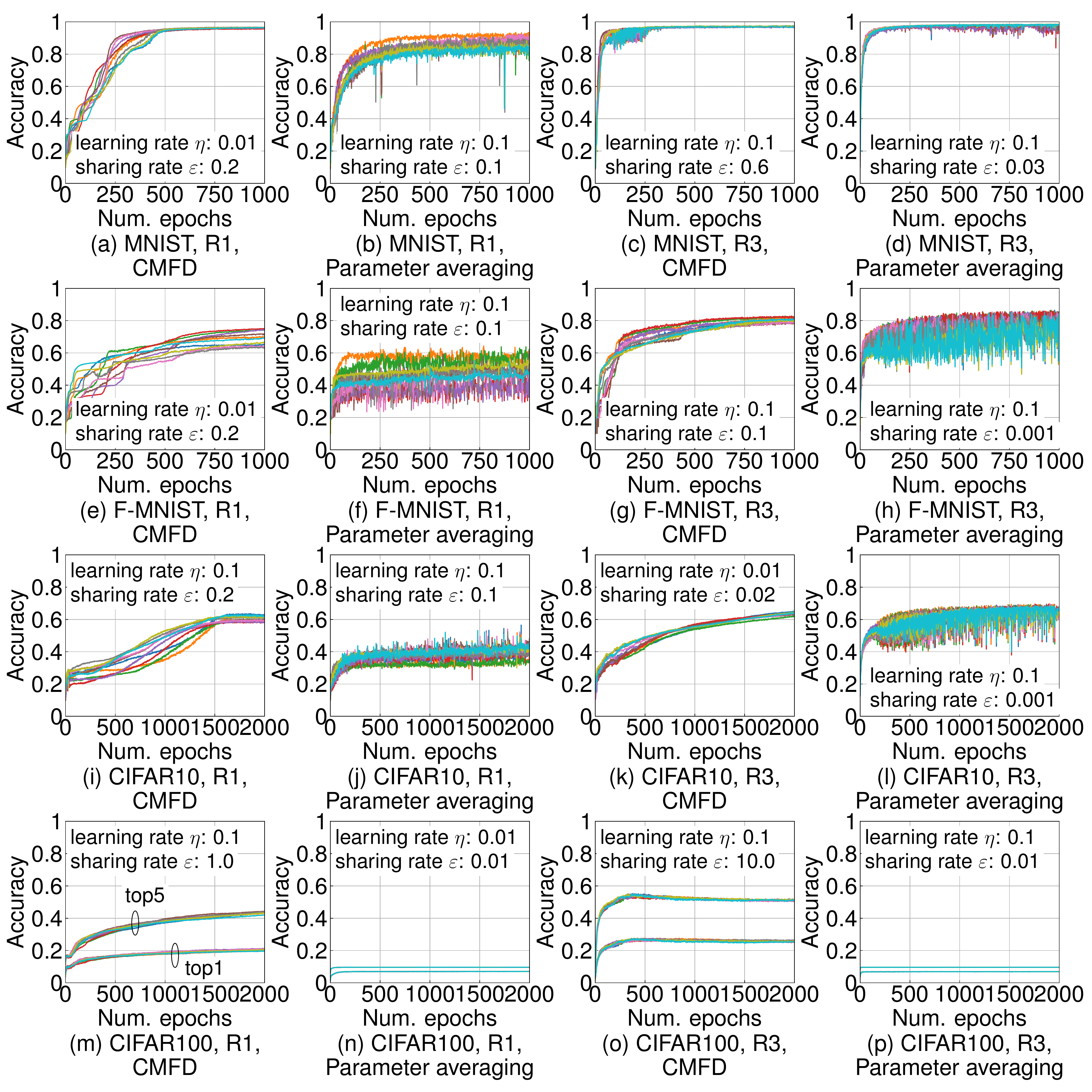}
\caption{Accuracy as a function of the number of epochs.
Each line represents the accuracy of each device.
The difference among devices with \gls{CMFD} is smaller compared with that of the parameter-averaging method.
Additionally, \gls{CMFD} achieves smaller fluctuation in time series.
Comparing network topologies R1 and R3, the strongly connected network R3 converges faster than R1.
}
\label{fig:eval_epoch2acc}
\end{figure}

Fig.~\ref{fig:eval_epoch2acc} shows the convergence performance evaluated for accuracy using MNIST, \gls{FMNIST}, CIFAR10, and CIFAR100.
The figures on the even columns show the performance of the parameter-averaging method as a baseline.
Each line in the figures indicates the accuracy of each device.
We manually tuned the sharing rate $\lrc$ for learning rates $\lr=0.01$ and $0.1$.
\Gls{CMFD} shows higher accuracy compared with the parameter-averaging method when the network is weakly connected.
Moreover, when learning CIFAR100, \gls{CMFD} outperforms the parameter-averaging method when the network is strongly connected.
These results suggest that \gls{CMFD} has the potential to learn complex datasets compared to the parameter-averaging method.
Regardless of which dataset was used,
comparing R1 and R3 confirms that
accuracy increased more rapidly when the network was strongly connected.
Unsurprisingly, the trained models of the ten devices got close faster when the network became dense.
Therefore, each device increased the prediction accuracy of unknown labels that the device lacked.
It also shows that the fluctuation range of the \gls{CMFD} is narrower than that of the parameter-averaging algorithm.

\newcommand{\labelacc}{\textsf{acc}\xspace}
\newcommand{\labeldiff}{\textsf{max-min}\xspace}
\newcommand{\labeldev}{\textsf{dev}\xspace}
\newcommand{\labeltopf}{\textsf{top5}\xspace}
\begin{table*}[!t]
\caption{Accuracy with various parameters}
\label{tbl:eval_accuracy}
\centering
\begin{tabular}{cccccccccccccccccc}
\toprule
\multicolumn{3}{c}{} & \multicolumn{7}{c}{\gls{CMFD}} & \multicolumn{7}{c}{Parameter averaging} & \gls{PDMM} \cite{niwa2020edge} \\
\midrule
\multicolumn{3}{c}{Learning rate $\lr$} & \multicolumn{3}{c}{0.01} & \multicolumn{3}{c}{0.1} & \multirow{2}{*}{\textbf{Best}} & \multicolumn{3}{c}{0.01} & \multicolumn{3}{c}{0.1} & \multirow{2}{*}{\textbf{Best}} & \multirow{2}{*}{} \\
\multicolumn{3}{c}{Sharing rate $\lrc$} & $0.01$ & $0.1$ & $1$ & $0.01$ & $0.1$ & $1$ & & $0.001$ & $0.01$ & $0.1$ & $0.001$ & $0.01$ & $0.1$ & & \\
\midrule
\multirow{6}{*}{\rotatebox{90}{F-MNIST}}
& \multirow{3}{*}{R1}
   & \labelacc  & 30.9 & 60.6 & 64.1 & 20.8 & 39.7 & 62.4 & \textbf{69.2} & 44.1 & 44.4 & 45.4 & 51.4 & 50.3 & 51.9 & \textbf{51.9} & \textbf{77.7} \\
 & & \labeldiff & 21.3 & 15.7 & 26.0 & 10.8 & 36.7 & 26.4 &         12.2  & 13.9 & 12.4 & 13.0 & 18.9 & 15.8 & 15.7 &         15.7  & 0.3  \\
 & & \labeldev  &  0.2 &  0.3 &  0.4 &  0.0 &  0.2 &  0.8 &          0.2  &  0.9 &  0.8 &  0.7 &  3.2 &  3.1 &  2.8 &          2.8  & 0.2  \\
\cmidrule{2-18}
& \multirow{3}{*}{R3}
   & \labelacc  & 60.6 & 77.2 & 62.8 & 38.9 & 80.6 & 62.8 & \textbf{80.6} & 67.3 & 65.7 & 65.9 & 82.1 & 80.9 & 74.6 & \textbf{82.1} & \textbf{76.1} \\
 & & \labeldiff & 13.7 &  5.1 &  9.3 & 37.7 &  3.8 & 11.3 &          3.8  & 12.6 & 18.4 & 12.5 &  9.5 &  6.9 & 18.5 &          9.5  & 0.3 \\
 & & \labeldev  &  0.3 &  0.3 &  0.7 &  0.3 &  0.3 &  0.5 &          0.3  &  2.0 &  1.7 &  1.3 &  5.8 &  5.6 &  6.9 &          5.8  & 0.2 \\
\midrule
\multirow{6}{*}{\rotatebox{90}{CIFAR10}}
& \multirow{3}{*}{R1}
   & \labelacc  & 38.8 & 46.0 & 41.3 & 37.4 & 59.2 & 39.1 & \textbf{60.3} & 30.1 & 30.1 & 30.4 & 39.8 & 39.9 & 40.2 & \textbf{40.2} & \textbf{51.7} \\
 & & \labeldiff & 10.9 &  9.3 &  8.3 & 13.7 &  4.9 & 13.8 &           4.5 & 10.3 &  9.5 &  7.0 & 10.5 & 10.3 & 14.9 &          14.9 & 0.4  \\
 & & \labeldev  &  0.2 &  0.5 &  0.3 &  0.2 &  0.2 &  0.3 &           0.2 &  0.8 &  0.7 &  0.7 &  1.5 &  1.3 &  1.7 &           1.7 & 0.1  \\
\cmidrule{2-18}
& \multirow{3}{*}{R3}
   & \labelacc  & 62.1 & 61.7 & 53.1 & 61.0 & 54.7 & 53.9 & \textbf{63.3} & 54.6 & 51.8 & 44.5 & 66.2 & 65.8 & 63.2 & \textbf{66.2} & \textbf{47.5} \\
 & & \labeldiff &  2.9 &  2.5 &  3.0 &  4.4 &  3.8 &  3.4 &           2.6 & 12.9 & 19.7 & 15.7 &  3.7 &  2.7 &  4.1 &           3.7 & 0.4  \\
 & & \labeldev  &  0.1 &  0.3 &  0.3 &  0.1 &  0.3 &  0.7 &           0.2 &  5.5 &  6.2 &  4.9 &  3.3 &  2.6 &  4.1 &           3.3 & 0.2  \\
\midrule
\multirow{8}{*}{\rotatebox{90}{CIFAR100}}
& \multirow{4}{*}{R1}
   & \labelacc  & 10.6 & 14.7 & 18.3 & 10.9 & 13.8 & 20.4 & \textbf{20.4} &  6.6 &  7.0 &  6.7 &  6.8 &  6.9 &  6.9 &  \textbf{7.0} & \textbf{17.0} \\
 & & \labeltopf & 16.8 & 27.7 & 39.0 & 16.9 & 25.2 & 43.5 & \textbf{43.5} &  9.5 &  9.6 &  9.5 &  9.6 &  9.6 &  9.6 &  \textbf{9.6} & \textbf{40.6} \\
 & & \labeldiff &  1.4 &  2.4 &  1.3 &  1.4 &  1.3 &  1.4 &           1.4 &  0.0 &  0.1 &  0.1 &  0.1 &  0.1 &  0.0 &           0.1 & 0.3  \\
 & & \labeldev  &  0.0 &  0.1 &  0.1 &  0.0 &  0.1 &  0.1 &           0.1 &  0.0 &  0.0 &  0.0 &  0.0 &  0.0 &  0.0 &           0.0 & 0.2  \\
\cmidrule{2-18}
& \multirow{4}{*}{R3}
   & \labelacc  & 10.7 & 19.5 & 25.1 & 10.5 & 19.4 & 23.8 & \textbf{25.8} &  6.7 &  6.9 &  6.6 &  6.8 &  6.9 &  6.8 &  \textbf{6.9} & \textbf{12.6} \\
 & & \labeltopf & 19.5 & 37.9 & 51.0 & 17.3 & 35.7 & 47.8 & \textbf{51.1} &  9.6 &  9.5 &  9.6 &  9.5 &  9.6 &  9.6 &  \textbf{9.6} & \textbf{32.6} \\
 & & \labeldiff &  1.3 &  1.0 &  0.8 &  1.5 &  1.7 &  1.2 &           1.4 &  0.1 &  0.1 &  0.0 &  0.0 &  0.0 &  0.0 &           0.0 & 0.4  \\
 & & \labeldev  &  0.1 &  0.1 &  0.1 &  0.0 &  0.1 &  0.1 &           0.1 &  0.0 &  0.0 &  0.0 &  0.0 &  0.0 &  0.0 &           0.0 & 0.2  \\
\bottomrule
\multicolumn{18}{l}{\labelacc\ (\%): Accuracy}\\
\multicolumn{18}{l}{\labeldiff\ (pp): Difference in the accuracies of the best and the worse devices}\\
\multicolumn{18}{l}{\labeldev\ (pp): Averaged standard deviation during the last 100 epochs}\\
\multicolumn{18}{l}{\labeltopf\ (\%): Top-5 accuracy}
\end{tabular}
\end{table*}

Table~\ref{tbl:eval_accuracy} lists the average accuracy and the other two metrics at 1000 epochs (\gls{FMNIST}) and 2000 epochs (CIFAR10 and CIFAR100) with various parameters.
Metric \labeldiff shows the difference in the accuracy of the best and the worse devices,
which is a metric of convergence among devices.
Label \labeldev shows the averaged standard deviation of each device's accuracy during the last 100 epochs.
Values of \labeldiff confirm that
when the learning rate and the sharing rate were well-tuned,
the accuracies of the ten devices
converged in narrower ranges with \gls{CMFD} than parameter averaging.
By comparing the deviations of the two methods, the fluctuation range of \gls{CMFD} is approximately one-tenth that of parameter averaging.
The difference in stability occurs because the parameter-averaging algorithm attempts to solve a non-convex optimization problem,
whereas \gls{CMFD} is designed to solve a convex optimization problem in a function space.
Because there are multiple local optima in a parameter space, when the parameters of devices tend toward different optima,
the parameter-updating directions calculated by parameter sharing and gradient descent can differ completely.
Conversely, the local prediction functions of devices draw close in a function space by \gls{CMFD},
which allows different parameters if they are located at the same point in a function space.
This difference in stability is significant, especially if the devices are weakly connected,
because the variance of the parameter-updating directions increases owing to the small number of adjacent devices
when using the parameter-averaging scheme.
In contrast, \gls{CMFD}, which aims to aggregate functions directly,
is not exposed to a major variance in the function-updating directions even with few adjacent devices.

Table~\ref{tbl:eval_accuracy} lists the evaluated results of \gls{PDMM} \cite{niwa2020edge} as the state-of-the-art decentralized \gls{FL}.
The communication frequency of \gls{PDMM} is set such that the number of transmissions is approximately same as \gls{CMFD} and parameter averaging
because \gls{PDMM} is an asynchronous algorithm, whereas \gls{CMFD} is a synchronous algorithm.
In this setting, the amount of transmitted data is twice as large as that of parameter averaging.
Although \gls{PDMM} achieves high convergence among devices and high accuracy even when devices are weakly connected,
\gls{CMFD} outperforms \gls{PDMM} in terms of accuracy on CIFAR10 and CIFAR100.
\Gls{CMFD} regards learning tasks as convex function optimization problems,
whereas \gls{PDMM} and parameter averaging attempt to solve non-convex optimization problems in the parameter space.
Therefore, \gls{CMFD} achieves higher accuracy for difficult tasks compared with the other methods.



\subsection{Heterogeneous environment}
\begin{figure}[!t]
\centering
\includegraphics[width=\figsizeMixedAll]{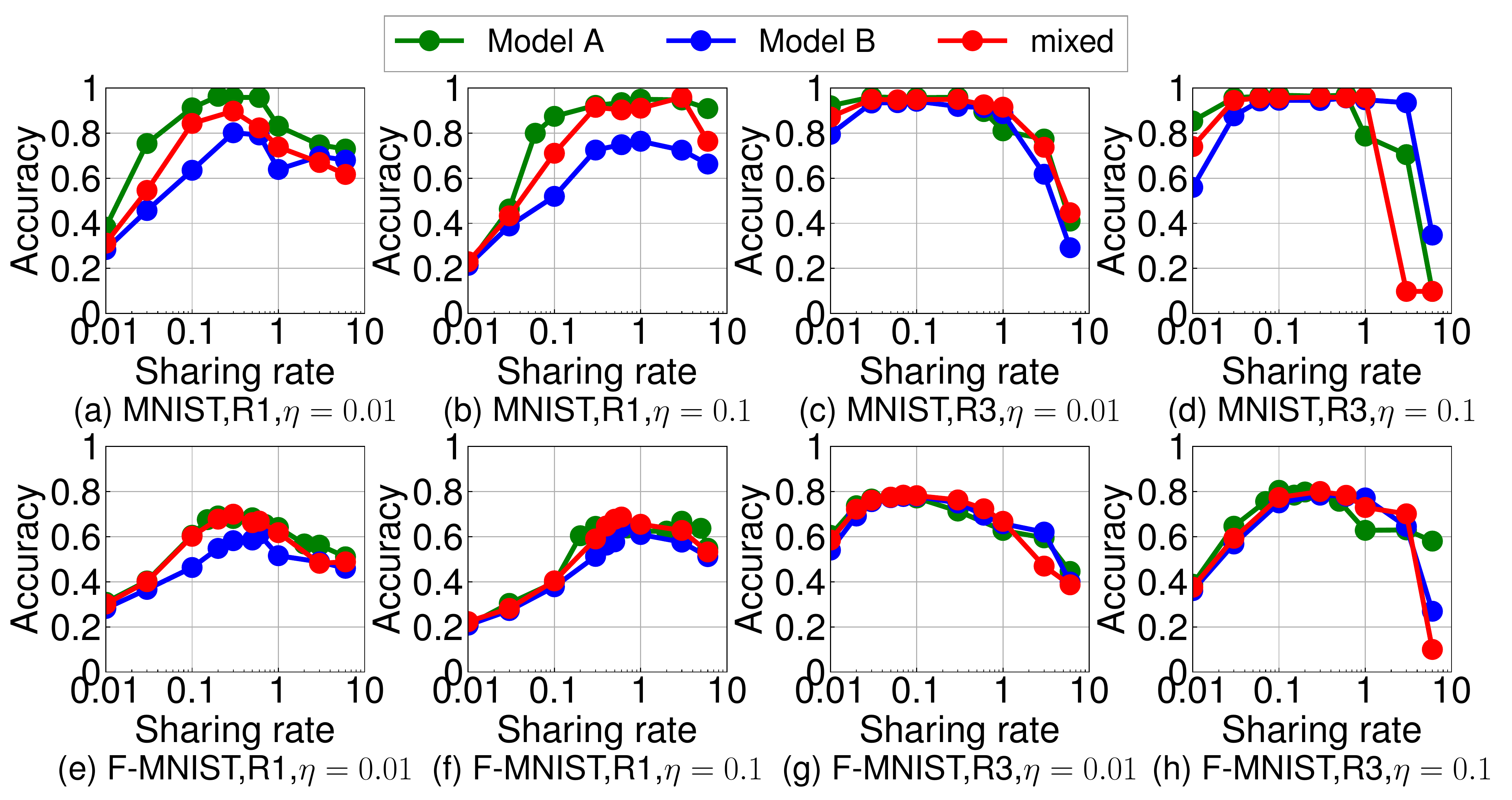}
\caption{Accuracy when different models are adopted.
Because Model A is more complex than Model B, Model A tends to achieve higher performance than Model B,
especially when the devices are weakly connected.
In a heterogeneous environment, devices using Model B achieved accuracy similar to those using Model A.
}
\label{fig:eval_mixedmodel}
\end{figure}

Fig.~\ref{fig:eval_mixedmodel} shows accuracy as a function of the sharing rates using different types of \gls{NN} models.
Model A's layer architecture was CNN($32, 5$)-CNN($64, 5$)-FC($512$)-FC($10$), which is identical to that used in the previous evaluation.
Model B was evaluated as a shallow \gls{NN} with CNN($8, 5$)-FC($32$)-FC($10$)-layer architecture.
The green and blue lines show the accuracy when all devices adopted models A and B, respectively.
The red line shows the accuracy when half of the ten devices adopted model A and the others adopted model B
considering a situation where the devices had different computational resources.
During the evaluation, the devices with an even-numbered index had rich resources and adopted model A,
and the others had poor resources and adopted shallow model B.
\Gls{CMFD} works even when devices adopt different \gls{NN} models,
whereas \gls{FL} algorithms using parameter averaging require the devices to adopt the same \gls{NN} model.
Because model B is shallow, its accuracy is lower than that of model A when the network topology is R1.
In contrast, when the connectivity increased, the accuracy of model B increased to the same level as that of model A.
This implies that the restrictions of local training perturbed the learning process, and the shallow model could not fulfill its potential.
When the two types of models were adopted by the different devices,
those that adopted model B achieved nearly the same performance as those using model A.
Such characteristics enable \gls{IoE} devices with different computational resources to cooperatively learn the same task.

\begin{figure}[!t]
\centering
\includegraphics[width=\figsizeMixedThree]{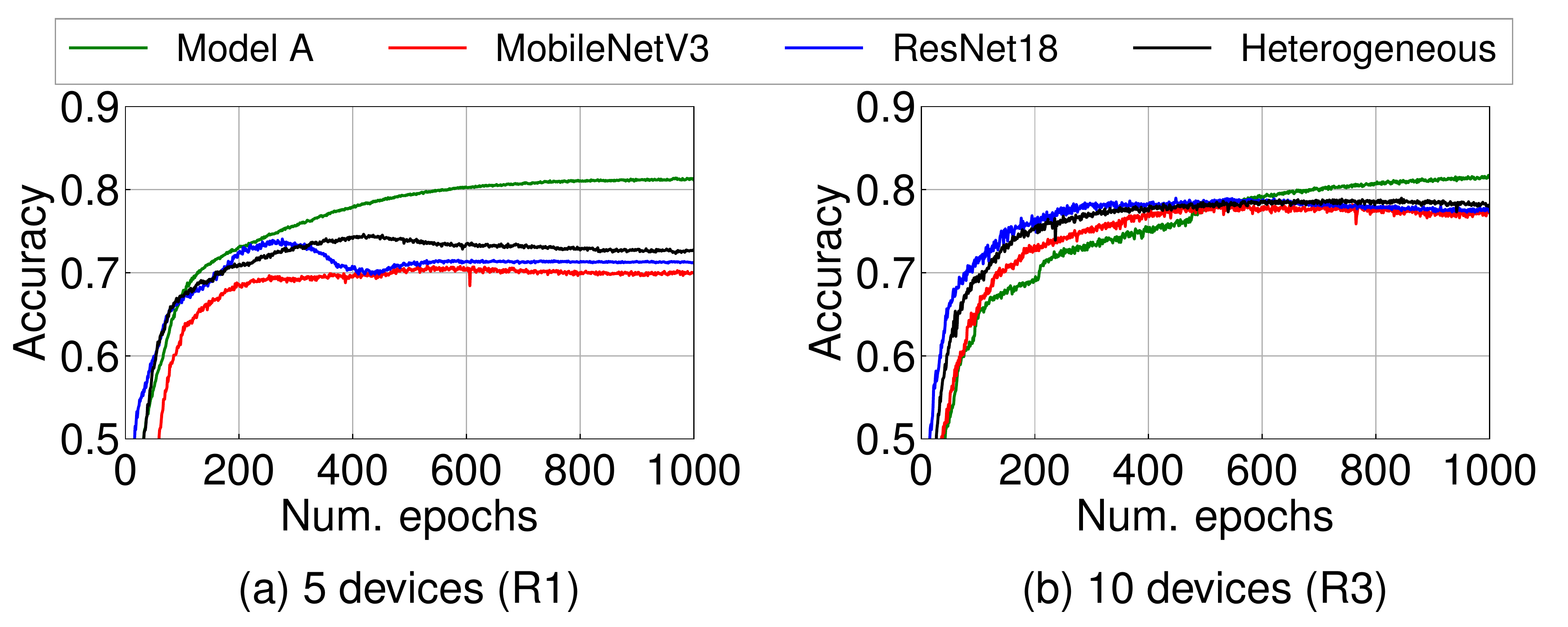}
\caption{
Accuracy as a function of the number of epochs when three types of models are adopted.
Green, red, and blue lines represent the accuracy for the homogenous cases where the devices use the same model architecture, i.e., Model A, MobileNetV3, or ResNet18.
Black lines represent the accuracy for the heterogeneous cases where the devices adopt different models.
Devices can optimize their local models even in a heterogeneous situation using \gls{CMFD}.
}
\label{fig:eval_mixedmodel_conv}
\end{figure}

We conducted another simulation using three types of models---model A, MobileNetV3 Small \cite{howard2019searching}, and ResNet18 \cite{he2016deep}---%
to verify whether \gls{CMFD} can learn more complex models in heterogeneous settings.
Fig.~\ref{fig:eval_mixedmodel_conv} shows the accuracy as a function of the number of epochs when five or ten devices learn \gls{FMNIST}.
Each line represents the average accuracy among all devices.
When the number of devices is five, each device possesses a dataset consisting of four categories,
and the number of local data samples is 2000.
The dropout rate was set to 0.1, and the other parameters are listed in Table~\ref{tbl:sim_param1}.
The input images were resized to $96\times 96$ before being input into MobileNetV3 and ResNet18
because these models are considerably deep, and the image sizes decrease to one pixel in the last layer without expansion.
We substituted layer normalization \cite{ba2016layer} for batch normalization \cite{ioffe2015batch} in MobileNetV3 and ResNet18
because batch normalization is not suitable for non-\gls{IID} datasets \cite{niwa2020edge}.
Fig.~\ref{fig:eval_mixedmodel_conv} shows that \gls{CMFD} can train models with non-\gls{IID} datasets in heterogeneous scenarios
even though the numbers of the layers of the models are widely different:
Model A has only four layers, while MobileNetV3 Small and ResNet18 have 16 and 18 layers, respectively.

\subsection{Comparison with various sharing rates}
\begin{figure}[!t]
\centering
\includegraphics[width=\figsizeRobustAll]{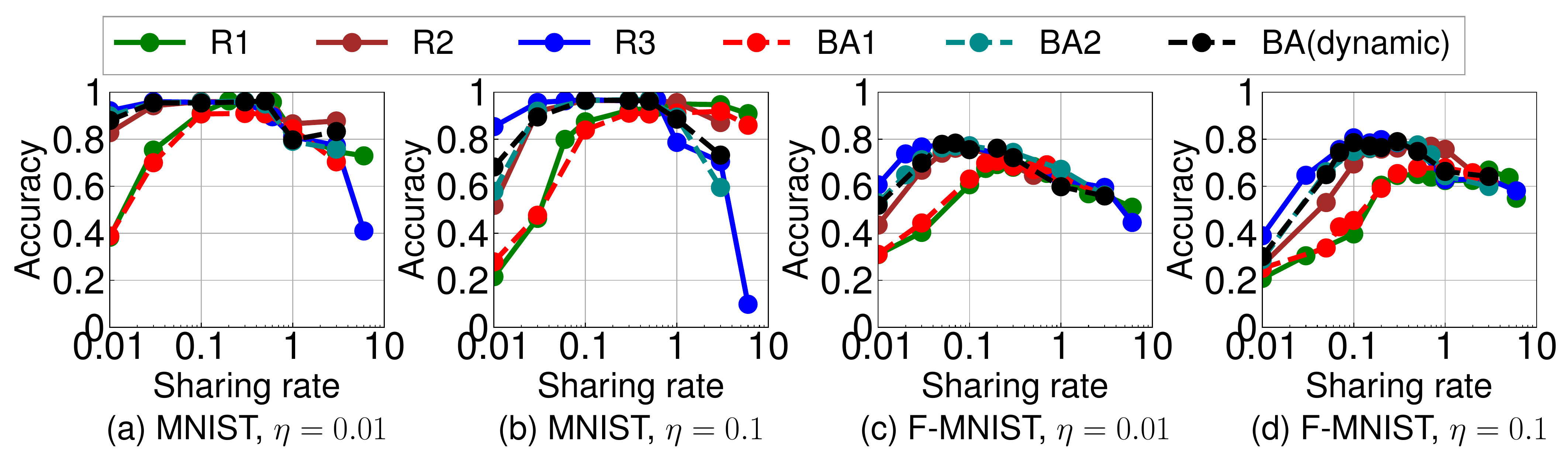}
\caption{Accuracy as a function of the sharing rate in various topologies.
When the network is strongly connected, accuracy increases.
There is apparently no difference between stable and unstable networks.
}
\label{fig:eval_sr2acc}
\end{figure}

\begin{table}[!t]
\caption{Connectivity of topologies}
\label{tbl:connectivity}
\centering
\begin{tabular}{cccccc}
\toprule
Topology & R1 & R2 & R3 & BA1 & BA2 \\
\midrule
Algebraic connectivity & 0.38 & 1.76 & 4.38 & 0.18 & 1.42 \\
Average of degrees & 2 & 4 & 6 & 1.8 & 4.2 \\
\bottomrule
\end{tabular}
\end{table}

Various topologies are compared in Fig.~\ref{fig:eval_sr2acc}, which shows the average accuracy as a function of the sharing rate.
The accuracy indicated by the label ``BA (dynamic)'' shows the performance when the network topology dynamically changed at each epoch
to evaluate situations where the mobile devices were in motion.
This shows that even if the topology changes dynamically, the accuracy converges to the same value.
Notably, these results also support that \gls{CMFD} is highly stable even when it is applied via wireless networks with considerable noise
because a dynamic topology emulates a situation with unstable links.
When comparing the topologies, the accuracy of R3 was slightly better than that of R2, BA2, and dynamic BA.
However, the accuracy of R1 and BA1 was poorer than that of the others.
These differences arise from the algebraic connectivity $\eigen_2$ in (\ref{eq:limDist}) and (\ref{eq:limBest}).
Table~\ref{tbl:connectivity} lists the algebraic connectivities and average degrees of the evaluated topologies.
Because the connectivity of R2, BA2, and dynamic BA are similar, they achieved similar performance.
Similarly, the accuracy of R1 and BA1 were similar because they had similar connectivity.

When evaluating ring topologies, adjacent devices have the same labels when evaluating ring networks,
whereas randomly selected labels are distributed when evaluating \gls{BA} networks.
Despite such data-distribution differences, they achieved similar accuracy if the connectivity of the topologies was nearly identical.
Therefore, it can be said that the data distribution similarity among adjacent devices does not affect accuracy.

\subsection{Scalability to the number of devices}
\begin{figure}[!t]
\centering
\includegraphics[width=\figsizeManyAgentsAll]{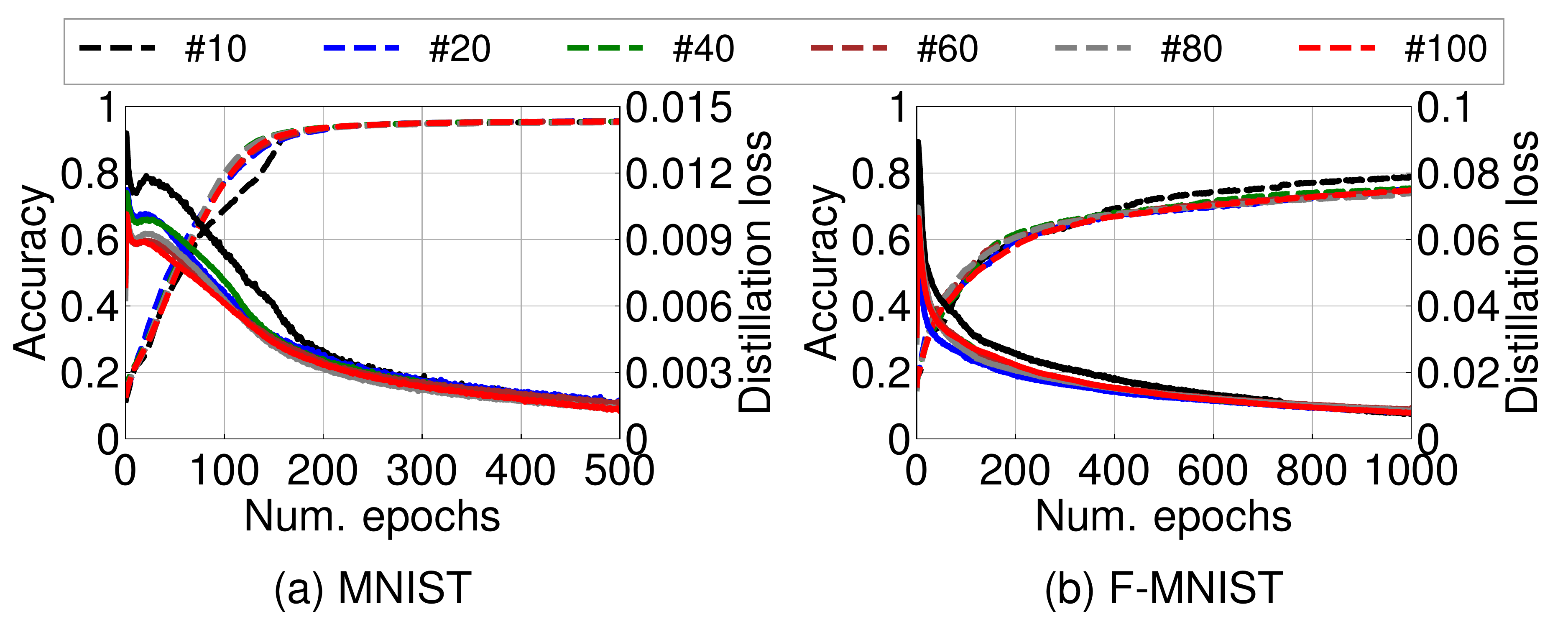}
\caption{Accuracy with various numbers of devices.
The dashed lines show the average accuracy among all devices, and the solid lines show the average distillation losses.
Both accuracy and distillation losses show similar values even when the number of devices increased
because the networks had similar algebraic connectivity regardless of the number of devices.
This means that the learning performance is scalable to \gls{FL} participants.
It is also shown that the local models converge to the same function
because distillation loss, which indicates the similarity of \gls{NN} models as functions, decreases with the number of epochs.
}
\label{fig:eval_num_devices}
\end{figure}
\begin{table}[!t]
\caption{Connectivity with various numbers of devices}
\label{tbl:connectivity_num}
\centering
\begin{tabular}{ccccccc}
\toprule
Num. devices & 10 & 20 & 40 & 60 & 80 & 100 \\
\midrule
Algebraic connectivity & 1.41 & 1.44 & 1.40 & 1.37 & 1.35 & 1.32 \\
Average of degrees & 4.2 & 5.1 & 5.55 & 5.7 & 5.775 & 5.82 \\
\bottomrule
\end{tabular}
\end{table}

Fig.~\ref{fig:eval_num_devices} shows the average accuracy and distillation losses for various numbers of devices.
The distillation loss is \gls{MSE} $c(\ws)$, defined in (\ref{eq:distillation_loss}).
This metric indicates the similarity of \gls{NN} models among adjacent devices
because \gls{MSE} is an empirical representation of the $\LSpc$ distance.
It is shown that distillation losses decrease with the number of epochs,
which means that the local models approach each other in the function space,
even though the parameters of the local models are not forced to converge by \gls{CMFD}.
Topologies varied at each epoch using the \gls{BA} algorithm,
but the number of all edges remained stable.
The connectivity for different numbers of devices are listed in Table~\ref{tbl:connectivity_num}.
The listed algebraic connectivity reflects the averages of 100 randomly generated graphs.
When evaluating MNIST, the learning and sharing rates were set to 0.01 and 0.1, respectively.
Both rates were set to 0.1 when evaluating \gls{FMNIST}.
Fig.~\ref{fig:eval_num_devices} shows that even when the number of devices increases,
the \gls{NN} models converge at nearly the same rate.
This is because, as shown in Table~\ref{tbl:connectivity_num}, the averages of algebraic connectivity are approximately the same regardless of the number of devices.
These results indicate that the proposed algorithm works with a large number of devices
as long as the connectivity does not change drastically.

\subsection{Amount of transferred data}
One of the advantages of using distillation is that the amount of shared data is smaller than that of parameter averaging schemes.
When using \gls{CMFD}, the amount of data transferred via each link per epoch is 40\,KB for $\Card{\sharedD}=1000$
and 400\,KB for $\Card{\sharedD}=10000$.
Here, all values are encoded in a four-byte-floating-point format.
When using the parameter averaging scheme, the amount of transferred data is 6.7\,MB with model A,
and 202\,KB for model B, respectively.
Because distillation-based methods require only to share the output values, when the input values are shared in advance,
the data volume can be smaller than the parameter-averaging schemes when \gls{NN} models have many parameters.


\section{Conclusion} \label{sec:conclusion}
We propose a model-free federated learning scheme for \gls{IoE} devices connected via multi-hop networks.
We developed a meta-algorithm for consensus-based optimization in a function space
that optimizes the loss function in a distributed manner.
The convergence of the meta-algorithm in a function space was mathematically analyzed using federated functions.
Federated functions represent the joint status of the entire system,
enabling us to obtain an upper bound of the distance between the local prediction functions.
Then, we propose a \gls{CMFD} as an implementation of the meta-algorithm,
which realizes function aggregation among adjacent devices by leveraging a distillation method.
We showed that the \gls{CMFD} achieved higher prediction accuracy and stability than a typical parameter-aggregation-based \gls{FL}
when a multi-hop network is weakly connected.

Our future work includes improving the algorithm by estimating the reliability of the local prediction functions.
When the training data is non-\gls{IID}, there is a training bias among the devices depending on the data distributions.
Although \gls{CMFD} does not consider the reliability of the prediction accuracy for each label,
the learning speed is expected to increase if weighted averaging is adopted when aggregating functions.
To calculate the optimal weight, we will develop a method of estimating how well the local prediction function is trained.

Extending \gls{CMFD} to non-convex optimization is an interesting study.
Although usual loss functionals are convex, non-convex functionals exist.
We believe that analyses of non-convex optimizations \cite{sun2020improving,lu2020decentralized} can be extended to function spaces
by leveraging Lemma~\ref{lemm:inducednorm}, which enables us to utilize the spectral graph theory in function spaces.

Another direction of future work is to apply a decentralized \gls{ADMM} to \gls{CMFD}
because it can achieve faster convergence compared with gradient-descent-based algorithms.
We believe \gls{ADMM} in a function space can be analyzed similarly to the analysis in Sec.~IV.
However, some challenges exist.
For example, devices have to calculate \gls{Frechet} gradients of complex functionals
depending on the local prediction functions of adjacent devices
to realize \gls{ADMM} in a function space.
This calculation is computationally expensive,
and devices must prepare multiple frameworks of \gls{NN} architectures in advance,
which decreases the flexibility of model selection in heterogeneous scenarios.
A novel approach should be developed to calculate gradients of \gls{NN} parameters that correspond to the \gls{Frechet} gradients
to improve practicality.

\appendices
\section{Toy model of \gls{CMFD}} \label{sec:app_toymodel}

\begin{figure}[!t]
\centering
\includegraphics[width=\figsizeToymodel, trim={0 40 0 0}, clip]{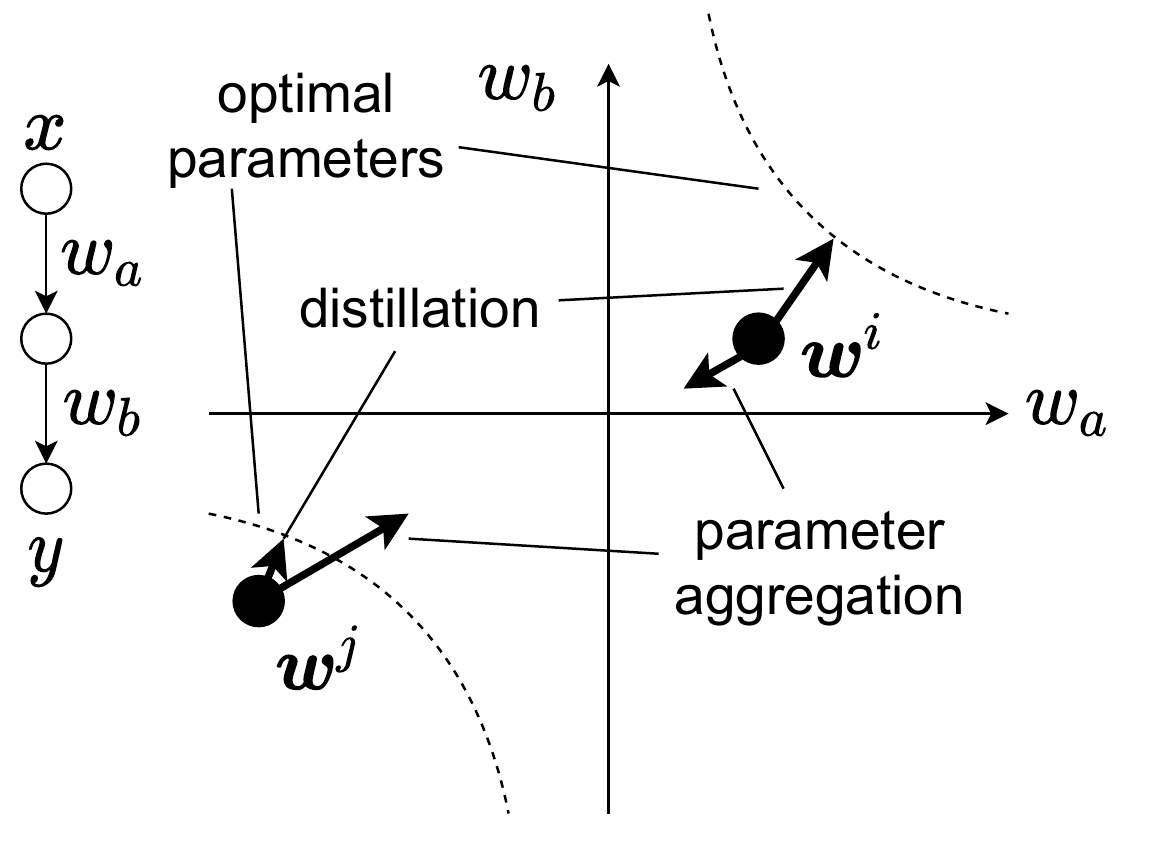}
\caption{\revhlm{(1-4) }{A toy model of \gls{CMFD}.
\Gls{CMFD} makes prediction models to approach each other in function space
allowing parameters to approach the different regions of the optimal parameters.
}
}
\label{fig:toymodel}
\end{figure}

\begin{revhl}{1-4}
Fig.~\ref{fig:toymodel} shows a simple example explaining
why a distillation-based algorithm is better than those trying to converge parameters.
Assume the ground truth model is defined by $\trueF(x) = x$,
and two devices $\#0$ and $\#1$ try to estimate $\trueF(x)$
using a three-layer perceptron with one hidden node and an identity activation function.
Then, the prediction model $f_i$ of device $i$ is represented by $f_i(x) = w_a^i w_b^i x$.
Let denote shared data by $x_n \in \sharedD$ and $j\coloneqq 1-i$.
\end{revhl}

\revhlnon{
In the setting, there are two optimal regions in a parameter space,
but these regions correspond to the same single point $\trueF$ in a function space.
Therefore, this optimization problem is non-convex in the parameter space
but convex in the function space.
}

\revhlnon{
Consider cases where
$w_a^i > 0, w_b^i > 0, w_a^j <0, w_b^j < 0, w_a^i w_b^i < 1 < w_a^j w_b^j$
as illustrated in Fig.~\ref{fig:toymodel}.
In such cases, when using an algorithm that forces parameter convergence,
$\bm{w}^i$ heads to approach $\bm{w}^j$ and goes away from the optimal region in the first quadrant.
}

\revhlnon{
In contrast, the parameters $\bm{w}^i$ approach the optimal region in the first quadrant with our algorithm
because the direction of updates by distillation is calculated as follows:
\begin{align}
  &- \frac{\partial}{\partial \bm{w}^i} \sum_{x\in \sharedD} \|f_i(x) - f_j(x)\|^2 \nonumber \\
  &\qquad \propto - (w_a^i w_b^i - w_a^j w_b^j) \begin{pmatrix}
    w_b^i \\
    w_a^i \\
  \end{pmatrix}. \nonumber
\end{align}
It is also notable that when $\bm{w}^i$ and $\bm{w}^j$ reach the different optimal regions,
our algorithm successfully stops,
while algorithms in parameter space continue to update parameters even though prediction models are identical with $\trueF$.
Therefore, our algorithm achieves better convergence than those converge parameters.
}

\section{Proof of Lemma~\ref{lemm:inducednorm}} \label{sec:app_inducednorm}
Let $a_i^m(\bm{x})$ and $a^m(\bm{x})$ represent
the $m$th element of the real vector $a_i(\bm{x})\in\OutSpc$
and a real vector defined as $\(a_1^m(\bm{x}),\ldots,a_n^m(\bm{x})\)^\Trans\in\Real^\NumU$,
respectively.
$\Dist{A\bm{a}}_{\FedSpc}^2$ can be derived as follows:
\begin{align}
\Dist{A\bm{a}}_{\FedSpc}^2
  &\mathop{\le}_{(\ref{eq:def_inducednorm})} \int_{\InSpc}\sum_{m=1}^\DimY \Dist{A}^2\Dist{a^m(\bm{x})}_{\Real^\NumU}^2 \dmu \nonumber \\
  &= \Dist{A}^2 \Dist{\bm{a}}_{\FedSpc}^2.
\end{align}
Now, we have $\Dist{A\bm{a}}_{\FedSpc} \le \Dist{A} \Dist{\bm{a}}_{\FedSpc}$.
Let $v_i$ be the $i$th element of $\argmax_{\bm{v}\in\Real^\NumU}\{\frac{\Dist{A\bm{v}}}{\Dist{\bm{v}}}\}$.
If $\hat{\bm{a}}$ is defined as $\hat{a}_i^m(\bm{x})\equiv v_i$ for all $m$ and $\bm{x}$,
$\hat{\bm{a}}$ satisfies $\Dist{A\hat{\bm{a}}}_{\FedSpc} = \Dist{A} \Dist{\hat{\bm{a}}}_{\FedSpc}$.
Therefore, $\Dist{A}$ is the induced 2-norm corresponding to $\FedSpc$.

\section{Proof of Theorem~\ref{th:convdist}} \label{sec:app_convdist}
Substituting (\ref{eq:meta_grad_all}) into (\ref{eq:meta_consensus_all}), the federated function $\bm{f}_{t+1}$ is derived as follows:
\begin{align}
  \bm{f}_{t+1}
     =& P \bm{f}_t - \lr_t P \bm{d}_t
     = P^{t} \bm{f}_1 - \sum_{\tau=1}^{t} \lr_{\tau} P^{t-\tau+1} \bm{d}_{\tau},
\end{align}
where $P$ is defined as $P\coloneqq I - \lrc \Laplacian$.
Since $\Ones \Laplacian=\bm{0}$, we have $\Ones P^t = \Ones (\forall t \in \mathbb{N})$, where $\bm{0}$ represents the zero matrix.
Therefore, the mean federated function $\FedMeanF_{t+1}$ is expressed as follows:
\begin{align}
  \FedMeanF_{t+1}
    = \frac{1}{\NumU} \Ones \bm{f}_{t+1}
    = \FedMeanF_1 - \sum_{\tau=1}^{t}\lr_\tau \FedMeanD_\tau. \label{eq:mean_f}
\end{align}
Now, distance between $\bm{f}_{t+1}$ and $\FedMeanF_{t+1}$ is bounded as follows:
\begin{align}
  &\Dist{\bm{f}_{t+1} - \FedMeanF_{t+1}}_{\FedSpc} \nonumber \\
      & \quad = \lv P^t \bm{f}_1
             - \sum_{\tau=1}^{t}\lr_\tau P^{t-\tau+1} \bm{d}_\tau
             - \FedMeanF_1 + \sum_{\tau=1}^{t}\lr_\tau \FedMeanD_\tau \rv_{\FedSpc} \nonumber \\
      &\quad \le \Dist{Q_t\bm{f}_1}_{\FedSpc} + \sum_{\tau=1}^{t}\lr_\tau \Dist{Q_{t-\tau+1}\bm{d}_{\tau}}_{\FedSpc}, \label{eq:distf}
\end{align}
where $Q_t$ is defined as $Q_t \coloneqq P^t - \frac{1}{\NumU}\Ones$.
Using Lemma~\ref{lemm:inducednorm}, (\ref{eq:distf}) can be derived as follows:
\begin{align}
  &\Dist{\bm{f}_{t+1} - \FedMeanF_{t+1}}_{\FedSpc} \nonumber \\
      &\quad \le \Dist{Q_t}\initF + \sum_{\tau=1}^{t}\lr_\tau \Dist{Q_{t-\tau+1}}\Dist{\bm{d}_{\tau}}_{\FedSpc}. \label{eq:distf2}
\end{align}
Lemma~\ref{lemm:inducednorm} enables
norms of federated functions $\Dist{Q_t\bm{f}_1}_{\FedSpc}$ and $\Dist{Q_{t-\tau+1}\bm{d}_{\tau}}_{\FedSpc}$
to be divided into the multiplication of norms of real matrices $\Dist{Q_t}$ and $\Dist{Q_{t-\tau+1}}$
and federated functions $\initF$ and $\Dist{\bm{d}_\tau}_{\FedSpc}$.
Therefore, we only have to discuss upper bounds of $\Dist{Q_t}$ and $\Dist{\bm{d}_{\tau}}_{\FedSpc}$
instead of complex $\Dist{Q_t\bm{f}_1}_{\FedSpc}$ and $\Dist{Q_{t-\tau+1}\bm{d}_{\tau}}_{\FedSpc}$.

First, we consider an upper bound of $\Dist{Q_t}$.
Using spectral decomposition, $\Laplacian$ can be written as $\Laplacian=\sum_{i=1}^\NumU \eigen_i u_i u_i^\Trans$,
where $\eigen_i$ and $u_i$ the $i$th smallest eigenvalues and the corresponding orthonormal eigenvectors of $\Laplacian$, respectively.
Since the network graph is connected, the smallest eigenvalue and the corresponding eigenvector satisfies
$\eigen_1=0, u_1=\frac{1}{\sqrt{\NumU}}\Ones[\NumU]$.
Now, we can obtain the eigenvalues of $Q_t$ as follows:
\begin{align}
  P   &= I - \lrc \sum_{i=1}^\NumU \eigen_i u_i u_i^\Trans
      = \sum_{i=1}^\NumU (1 - \lrc \eigen_i) u_i u_i^\Trans, \\
  Q_t &= \sum_{i=1}^\NumU (1 - \lrc \eigen_i)^t u_i u_i^\Trans - \frac{1}{\NumU} \Ones
      = \sum_{i=2}^\NumU \eigenQ_i^t u_i u_i^\Trans,
\end{align}
where $\eigenQ_i$ denotes $1-\lrc \eigen_i$.
Thus, the eigenvalues of $Q_t$ become $0$ and $\eigenQ_i^t$ for $i\ge2$.
Considering that $\Laplacian$ is symmetric, $Q_t$ is also symmetric.
Therefore, induced norm $\Dist{Q_t}$ is equal to the largest eigenvalues of $Q_t$ in absolute.
Since the network graph is connected, $\eigen_i$ satisfies $0 < \eigen_i\le 2\MaxD$ for $i\ge 2$.
Therefore, if $\lrc$ satisfies $0 < \lrc \le \frac{1}{2\MaxD}$, we have $0\le \eigenQ_i < 1$.
Now, we have $\Dist{Q_t}=\max\{0, \eigenQ_2^t, \dots, \eigenQ_{\NumU}^t\} = \convRate^t$ because $\eigenQ_i$ monotonically decreases.

Next, we derive an upper bound of $\Dist{\bm{d}_{t}}_{\FedSpc}$.
Let $d \in \subDifLmui[f]$ denote a \gls{Frechet} subgradient of $\Lmui(f)$.
Using Lipschitz constant $\Lip{i}$ of $\Lmui(f)$ and the supremum $\maxRn$ of $\rn_i$,
an upper bound of $\Dist{d \rn_i}_{\LSpc}$ is obtained as follows:
\begin{align}
  \Dist{d \rn_i}_{\LSpc}
    &= \sqrt{\int_{\InSpc} \Dist{d}_{\OutSpc}^2 (\rn_i)^2 \dmu}
    = \sqrt{\int_{\InSpc} \Dist{d}_{\OutSpc}^2 \rn_i \dmui} \nonumber \\
    &\le \sqrt{\int_{\InSpc} \Dist{d}_{\OutSpc}^2 \maxRn \dmui}
    \le \sqrt{\maxRn} \Lip{i}. \label{eq:bound_dq}
\end{align}
Now, an upper bound of $\Dist{\bm{d}_{t}}_{\FedSpc}$ can be obtained as follows:
\begin{align}
  \Dist{\bm{d}_{t}}_{\FedSpc}
    &= \sqrt{\sum_i^{\NumU} \Dist{d_i^t \rn_i}_{\LSpc}^2}
    \le \sqrt{\NumU} \maxLip,
\end{align}
where $\maxLip$ is defined as $\maxLip\coloneqq \max_{i} \left\{\sqrt{\maxRn} \Lip{i}\right\}$.
Now, an upper bound of $\Dist{\bm{f}_{t+1} - \FedMeanF_{t+1}}_{\FedSpc}$ is obtained as follows:
\begin{align}
  &\Dist{\bm{f}_{t+1} - \FedMeanF_{t+1}}_{\FedSpc}
      \le \initF \convRate^t + \sqrt{\NumU} \maxLip \sum_{\tau=1}^{t} \lr_\tau \convRate^{t-\tau+1}.
\end{align}
Thus, we have the upper bound of the root-mean square distance $\MeanDist_t$ as follows:
\begin{align}
  \MeanDist_t
    &\le \frac{1}{\sqrt{n}} \initF \convRate^{t-1} + \maxLip\sum_{\tau=1}^{t-1} \lr_{\tau} \convRate^{t-\tau}.
\end{align}

\section{Proof of Theorem~\ref{th:convopt}} \label{sec:app_convopt}
Since the average of $f_i^t$ is written as $\MeanF_{t+1}=\MeanF_{t} - \lr_t \MeanD_{t}$,
we have
\begin{align}
  &\Dist{\MeanF_{t+1} - \trueF}_{\LSpc}^2
    = \Dist{\MeanF_{t} - \trueF}_{\LSpc}^2 \nonumber \\
    &\qquad \mbox{} - 2 \lr_t \inner{\MeanD_t}{\MeanF_t - \trueF}{\LSpc} + \lr_t^2 \Dist{\MeanD_t}_{\LSpc}^2. \label{eq:distf_true}
\end{align}
The second term of the \gls{RHS} of (\ref{eq:distf_true}) is derived as follows:
\begin{align}
  &- \inner{\MeanD_t}{\MeanF_t - \trueF}{\LSpc}
         = \frac{1}{\NumU} \sum_{i}^{\NumU} \inner{d_i^t \rn_i}{\trueF - \MeanF_t}{\LSpc} \nonumber \\
  &\quad = \frac{1}{\NumU} \sum_{i}^{\NumU}
            \[\inner{d_i^t \rn_i}{\trueF - f_i^t}{\LSpc}
            + \inner{d_i^t \rn_i}{f_i^t - \MeanF_t}{\LSpc}\]. \label{eq:ineqA}
\end{align}
Using Cauchy--Schwarz inequality, we have
\begin{align}
  \inner{d_i^t \rn_i}{f_i^t - \MeanF_t}{\LSpc}
    &\le \Dist{d_i^t \rn_i}_{\LSpc} \Dist{f_i^t - \MeanF_t}_{\LSpc} \nonumber \\
    &\mathop{\le}_{(\ref{eq:bound_dq})} \maxLip \Dist{f_i^t - \MeanF_t}_{\LSpc}. \label{eq:ineqB}
\end{align}
Using Radon-Nikodym derivative $\rn_i=\frac{\dmui}{\dmu}$, (\ref{eq:subdifferential}) can be rewritten as follows:
\begin{align}
  \forall h\in\LSpc[\mu_i], 
  \inner{d_i^t \rn_i}{h-f_i^t}{\LSpc} \le \Lmui(h) - \Lmui(f_i^t).
\end{align}
Choosing $\trueF$ as $h$, we have
\begin{align}
  &\inner{d_i^t \rn_i}{\trueF-f_i^t}{\LSpc}
         \le \Lmui(\trueF) - \Lmui(f_i^t) \nonumber \\
  &\quad \le \Lmui(\trueF) - \Lmui(\MeanF_t) + \Lmui(\MeanF_t) - \Lmui(f_i^t). \label{eq:ineqC}
\end{align}
Here, let $\gradAvg$ denote a \gls{Frechet} subgradient of $\Lmui(\MeanF_t)$, 
then the $\gradAvg$ satisfies the following inequality:
\begin{align}
  \forall h, \quad \inner{\gradAvg}{h-\MeanF_t}{\LSpc[\mu_i]} \le \Lmui(h) - \Lmui(\MeanF_t).
\end{align}
If we choose $f_i^t$ as $h$, we have
\begin{align}
  &\Lmui(\MeanF_t) - \Lmui(f_i^t)
    \le \inner{\gradAvg}{\MeanF_t - f_i^t}{\LSpc[\mu_i]} \nonumber \\
    &= \inner{\gradAvg \rn_i}{\MeanF_t - f_i^t}{\LSpc}
    \le \Dist{\gradAvg \rn_i}_{\LSpc} \Dist{\MeanF_t - f_i^t}_{\LSpc} \nonumber \\
    &\mathop{\le}_{(\ref{eq:bound_dq})} \maxLip \Dist{f_i^t - \MeanF_t}_{\LSpc}
    \le \sqrt{\NumU} \maxLip \distAvg_t. \label{eq:ineqD}
\end{align}
In the above inequality, we used $\Dist{f_i^t -\MeanF_t}_{\LSpc} \le \Dist{\bm{f}_t - \FedMeanF_t}_{\FedSpc}$.
Substituting (\ref{eq:ineqB}), (\ref{eq:ineqC}), and (\ref{eq:ineqD}) into (\ref{eq:ineqA}),
and using $\mu = \frac{1}{\NumU} \sum_{i\in\Users} \mu_i$,
we obtain the following inequality:
\begin{align}
  &- \inner{\MeanD_t}{\MeanF_t - \trueF}{\LSpc} \nonumber \\
  &\quad \le \frac{1}{\NumU} \sum_{i}^{\NumU}
            \[ \Lmui(\trueF) - \Lmui(\MeanF_t) + 2 \sqrt{\NumU} \maxLip \distAvg_t\] \nonumber \\
  &\quad \le \Lmu(\trueF) - \Lmu(\MeanF_t) + 2 \sqrt{\NumU} \maxLip \distAvg_t. \label{eq:ineqE}
\end{align}
Now, (\ref{eq:distf_true}) can be derived as follow:
\begin{align}
  &\Dist{\MeanF_{t+1} - \trueF}_{\LSpc}^2
         \le \Dist{\MeanF_{t} - \trueF}_{\LSpc}^2 + \lr_t^2 \maxLip \nonumber \\
  &\qquad + 2 \lr_t \[\Lmu(\trueF) - \Lmu(\MeanF_t) + 2 \sqrt{\NumU} \maxLip \distAvg_{t}\] \nonumber \\
  &\quad \le \Dist{\MeanF_{2} - \trueF}_{\LSpc}^2
         + \maxLip \sum_{\tau=2}^{t}\lr_{\tau}^2
         + 4 \sqrt{\NumU} \maxLip \sum_{\tau=2}^{t} \lr_{\tau} \distAvg_{\tau} \nonumber \\
  &\qquad + 2 \sum_{\tau=2}^{t} \lr_{\tau} \[\Lmu(\trueF) - \Lmu(\MeanF_{\tau})\].
\end{align}
Using $\Dist{\MeanF_{t+1} - \trueF}_{\LSpc}^2 \le 0$ and $\initDist \coloneqq \Dist{\MeanF_{2} - \trueF}_{\LSpc}^2$,
we have
\begin{align}
  &\Lmu(\bestF) - \Lmu(\trueF) = \min_{\tau=1,\dots,t}\{\Lmu(\MeanF_\tau)\} - \Lmu(\trueF) \nonumber \\
  &\quad \le \frac{
                \initDist + \maxLip \sum_{\tau=2}^{t}\lr_{\tau}^2
                + 4 \sqrt{\NumU} \maxLip \sum_{\tau=2}^{t} \lr_{\tau} \distAvg_{\tau}}
              {2 \sum_{\tau=2}^{t} \lr_{\tau} }. \label{eq:best}
\end{align}
An upper bound of $\sum_{\tau=2}^{t} \lr_{\tau} \distAvg_{\tau}$ is obtained as follows:
\begin{align}
  \sum_{\tau=2}^{t} \lr_{\tau} \distAvg_{\tau}
    =&  \frac{1}{\sqrt{\NumU}}\initF \sum_{\tau=2}^t \lr_{\tau} \convRate^{\tau-1} \nonumber \\
     &+ \maxLip \sum_{\tau=2}^{t} \sum_{\tau'=1}^{\tau-1} \lr_{\tau} \lr_{\tau'} \convRate^{\tau-\tau'}.
\end{align}
Using the monotonicity of $\lr_t$, the second term of the \gls{RHS} is upper bounded as follows:
\begin{align}
  &\sum_{\tau=2}^{t} \sum_{\tau'=1}^{\tau-1} \lr_{\tau} \lr_{\tau'} \convRate^{\tau-\tau'}
     = \sum_{\tau'=1}^{t-1}\sum_{\tau=1}^{t-\tau'} \lr_{\tau+\tau'} \lr_{\tau} \convRate^{\tau'} \nonumber \\
  &\quad \le \sum_{\tau'=1}^{t-1}\sum_{\tau=1}^{t-1} \lr_{\tau}^2 \convRate^{\tau'}
    = \frac{\convRate(1-\convRate^{t-1})}{1-\convRate} \sum_{\tau=1}^{t-1}\lr_{\tau}^2.
\end{align}
Thus, we have
\begin{align}
  \sum_{\tau=2}^{t} \lr_{\tau} \distAvg_{\tau}
    &\le \frac{\convRate(1-\convRate^{t-1})}{1-\convRate}\(\frac{\lr_1}{\sqrt{\NumU}} \initF + \maxLip \sum_{\tau=1}^{t-1}\lr_{\tau}^2\).
\end{align}
Now, (\ref{eq:best}) can be derived as follows:
\begin{align}
  &\Lmu(\bestF) - \Lmu(\trueF)
         \le \frac{1}{2 \sum_{\tau=2}^{t} \lr_{\tau}}
             \[ \initDist + \maxLip \sum_{\tau=2}^{t}\lr_{\tau}^2 \right. \nonumber \\
  &\qquad    \left. + \Coeff \(1-\convRate^{t-1}\) \(\lr_1 \initF + \sqrt{\NumU} \maxLip \sum_{\tau=1}^{t-1}\lr_{\tau}^2\)\], \nonumber
\end{align}
where $\Coeff$ is defined as $\Coeff\coloneqq \frac{4 \maxLip \convRate}{1-\convRate}$.

\section*{Acknowledgment}
This work was supported in part by JSPS KAKENHI Grant Number JP21K17734 and JST PRESTO Grant Number JPMJPR2035.

\ifCLASSOPTIONcaptionsoff
  \newpage
\fi


\bibliographystyle{IEEEtran}
\bibliography{IEEEabrv,main}

\begin{thebibliography}{10}
\providecommand{\url}[1]{#1}
\csname url@samestyle\endcsname
\providecommand{\newblock}{\relax}
\providecommand{\bibinfo}[2]{#2}
\providecommand{\BIBentrySTDinterwordspacing}{\spaceskip=0pt\relax}
\providecommand{\BIBentryALTinterwordstretchfactor}{4}
\providecommand{\BIBentryALTinterwordspacing}{\spaceskip=\fontdimen2\font plus
\BIBentryALTinterwordstretchfactor\fontdimen3\font minus
  \fontdimen4\font\relax}
\providecommand{\BIBforeignlanguage}[2]{{%
\expandafter\ifx\csname l@#1\endcsname\relax
\typeout{** WARNING: IEEEtran.bst: No hyphenation pattern has been}%
\typeout{** loaded for the language `#1'. Using the pattern for}%
\typeout{** the default language instead.}%
\else
\language=\csname l@#1\endcsname
\fi
#2}}
\providecommand{\BIBdecl}{\relax}
\BIBdecl

\bibitem{mcmahan2016communication}
B.~McMahan, E.~Moore, D.~Ramage, S.~Hampson, and B.~A. y~Arcas,
  ``Communication-efficient learning of deep networks from decentralized
  data,'' in \emph{Proc.\ 20th Int. Conf. Artificial Intelligence and
  Statistics (AISTATS)}, Fort Lauderdale, FL, USA, Apr. 2017, pp. 1273--1282.

\bibitem{kairouz2019advances}
P.~Kairouz \emph{et~al.}, ``Advances and open problems in federated learning,''
  \emph{Foundations and Trends{\textregistered} in Machine Learning}, vol.~14,
  no. 1--2, pp. 1--210, Jun. 2021.

\bibitem{savazzi2020federated}
S.~Savazzi, M.~Nicoli, and V.~Rampa, ``Federated learning with cooperating
  devices: A consensus approach for massive {IoT} networks,'' \emph{IEEE
  Internet Things J.}, vol.~7, no.~5, pp. 4641--4654, May 2020.

\bibitem{lalitha2019peer}
A.~Lalitha, O.~C. Kilinc, T.~Javidi, and F.~Koushanfar, ``Peer-to-peer
  federated learning on graphs,'' \emph{arXiv preprint arXiv:1901.11173}, Jan.
  2019.

\bibitem{lian2017can}
X.~Lian, C.~Zhang, H.~Zhang, C.-J. Hsieh, W.~Zhang, and J.~Liu, ``Can
  decentralized algorithms outperform centralized algorithms? a case study for
  decentralized parallel stochastic gradient descent,'' in \emph{Proc.\ 31st
  Conf. Neural Information Processing Systems (NeurIPS)}, Long Beach, CA, USA,
  Dec. 2017, pp. 5330--5340.

\bibitem{niwa2020edge}
K.~Niwa, N.~Harada, G.~Zhang, and W.~B. Kleijn, ``Edge-consensus learning: Deep
  learning on {P2P} networks with nonhomogeneous data,'' in \emph{Proc.\ 26th
  {ACM} {SIGKDD} Int. Conf. Knowledge Discovery \& Data Mining}, Virtual
  Conference, Aug. 2020, pp. 668--678.

\bibitem{sato2020network}
K.~Sato, Y.~Satoh, and D.~Sugimura, ``Network-density-controlled decentralized
  parallel stochastic gradient descent in wireless systems,'' in \emph{Proc.\
  IEEE Int. Conf. Commun. (ICC)}, Virtual Conference, Jun. 2020.

\bibitem{oh2020mix2fld}
S.~Oh, J.~Park, E.~Jeong, H.~Kim, M.~Bennis, and S.-L. Kim, ``{Mix2FLD}:
  downlink federated learning after uplink federated distillation with two-way
  mixup,'' \emph{IEEE Commun. Lett.}, vol.~24, no.~10, pp. 2211--2215, Jun.
  2020.

\bibitem{ahn2019wireless}
J.-H. Ahn, O.~Simeone, and J.~Kang, ``Wireless federated distillation for
  distributed edge learning with heterogeneous data,'' in \emph{Proc.\ 30th
  Annual International Symposium on Personal, Indoor and Mobile Radio
  Communications (PIMRC)}, Istanbul, Turkey, Nov. 2019, pp. 1--6.

\bibitem{itahara2021distillation}
S.~Itahara, T.~Nishio, Y.~Koda, M.~Morikura, and K.~Yamamoto,
  ``Distillation-based semi-supervised federated learning for
  communication-efficient collaborative training with non-{IID} private data,''
  \emph{IEEE Trans. Mobile Comput.}, pp. 1--15, Mar. 2021.

\bibitem{jeong2018communication}
E.~Jeong, S.~Oh, H.~Kim, J.~Park, M.~Bennis, and S.-L. Kim,
  ``Communication-efficient on-device machine learning: Federated distillation
  and augmentation under non-{IID} private data,'' \emph{arXiv preprint
  arXiv:1811.11479}, Nov. 2018.

\bibitem{chang2019cronus}
H.~Chang, V.~Shejwalkar, R.~Shokri, and A.~Houmansadr, ``Cronus: Robust and
  heterogeneous collaborative learning with black-box knowledge transfer,''
  \emph{arXiv preprint arXiv:1912.11279}, Dec. 2019.

\bibitem{li2019fedmd}
D.~Li and J.~Wang, ``{FedMD}: Heterogenous federated learning via model
  distillation,'' \emph{arXiv preprint arXiv:1910.03581}, Oct. 2019.

\bibitem{lin2020ensemble}
T.~Lin, L.~Kong, S.~U. Stich, and M.~Jaggi, ``Ensemble distillation for robust
  model fusion in federated learning,'' in \emph{Proc.\ 33rt Conf. Neural
  Information Processing Systems (NeurIPS)}, vol.~33, Virtual Conference, Dec.
  2020, pp. 2351--2363.

\bibitem{jeong2019multi}
E.~Jeong, S.~Oh, J.~Park, H.~Kim, M.~Bennis, and S.-L. Kim, ``Multi-hop
  federated private data augmentation with sample compression,'' \emph{arXiv
  preprint arXiv:1907.06426}, Jul. 2019.

\bibitem{anil2018large}
R.~Anil, G.~Pereyra, A.~Passos, R.~Ormandi, G.~E. Dahl, and G.~E. Hinton,
  ``Large scale distributed neural network training through online
  distillation,'' \emph{arXiv preprint arXiv:1804.03235}, Apr. 2018.

\bibitem{zhang2018deep}
Y.~Zhang, T.~Xiang, T.~M. Hospedales, and H.~Lu, ``Deep mutual learning,'' in
  \emph{Proc.\ 2018 IEEE/CVF Conf. Computer Vision and Pattern Recognition
  (CVPR)}.\hskip 1em plus 0.5em minus 0.4em\relax IEEE, 2018, pp. 4320--4328.

\bibitem{zhao2018federated}
Y.~Zhao, M.~Li, L.~Lai, N.~Suda, D.~Civin, and V.~Chandra, ``Federated learning
  with non-{IID} data,'' \emph{arXiv preprint arXiv:1806.00582}, Jun. 2018.

\bibitem{sattler2020robust}
F.~Sattler, S.~Wiedemann, K.-R. M{\"u}ller, and W.~Samek, ``Robust and
  communication-efficient federated learning from non-i.i.d. data,'' \emph{IEEE
  Trans. Neural Netw. Learn. Syst.}, vol.~31, no.~9, pp. 3400--3413, Sep. 2020.

\bibitem{li2020convergence}
X.~Li, K.~Huang, W.~Yang, S.~Wang, and Z.~Zhang, ``On the convergence of
  {FedAvg} on non-{IID} data,'' in \emph{Proc.\ of 7th Int. Conf. Learning
  Representations (ICLR)}, Online Conference, Apr. 2020.

\bibitem{mason1999boosting}
L.~Mason, J.~Baxter, P.~Bartlett, and M.~Frean, ``Boosting algorithms as
  gradient descent in function space,'' in \emph{Proc.\ 12th Int. Conf. Neural
  Information Processing Systems (NIPS)}, Cambridge, MA, USA, Nov. 1999, pp.
  512--518.

\bibitem{koppel2018decentralized}
A.~Koppel, S.~Paternain, C.~Richard, and A.~Ribeiro, ``Decentralized online
  learning with kernels,'' \emph{IEEE Trans. Signal Process.}, vol.~66, no.~12,
  pp. 3240--3255, Jun. 2018.

\bibitem{bucilua2006model}
C.~Buciluǎ, R.~Caruana, and A.~Niculescu-Mizil, ``Model compression,'' in
  \emph{Proc.\ 12th Int. Conf. Knowledge discovery and data mining (SIGKDD)},
  Philadelphia, USA, Aug. 2006, pp. 535--541.

\bibitem{park2019wireless}
J.~Park, S.~Samarakoon, M.~Bennis, , and M.~Debbah, ``Wireless network
  intelligence at the edge,'' \emph{Proc.\ IEEE}, vol. 107, no.~11, pp.
  2204--2239, Nov. 2019.

\bibitem{park2019distilling}
J.~Park, S.~Wang, A.~Elgabli, S.~Oh, E.~Jeong, H.~Cha, H.~Kim, S.-L. Kim, and
  M.~Bennis, ``Distilling on-device intelligence at the network edge,''
  \emph{arXiv preprint arXiv:1908.05895}, Aug. 2019.

\bibitem{nedic2009distributed}
A.~Nedic and A.~Ozdaglar, ``Distributed subgradient methods for multi-agent
  optimization,'' \emph{IEEE Trans. Autom. Control}, vol.~54, no.~1, pp.
  48--61, Jan. 2009.

\bibitem{johansson2010randomized}
B.~Johansson, M.~Rabi, and M.~Johansson, ``A randomized incremental subgradient
  method for distributed optimization in networked systems,'' \emph{SIAM J.
  Optimization}, vol.~20, no.~3, pp. 1157--1170, Aug. 2009.

\bibitem{colin2016gossip}
I.~Colin, A.~Bellet, J.~Salmon, and S.~Cl{\'e}men{\c{c}}on, ``Gossip dual
  averaging for decentralized optimization of pairwise functions,'' in
  \emph{Proc.\ 33rd Int. Conf. Machine Learning (ICML)}, New York, USA, Jun.
  2016, pp. 1388--1396.

\bibitem{elgabli2020gadmm}
A.~Elgabli, J.~Park, A.~S. Bedi, M.~Bennis, and V.~Aggarwal, ``{GADMM}: Fast
  and communication efficient framework for distributed machine learning,''
  \emph{Journal of Machine Learning Research}, vol.~21, no.~76, pp. 1--39, Mar.
  2020.

\bibitem{van2014renyi}
T.~Van~Erven and P.~Harremos, ``R{\'e}nyi divergence and {Kullback-Leibler}
  divergence,'' \emph{IEEE Trans. Inf. Theory}, vol.~60, no.~7, pp. 3797--3820,
  Jul. 2014.

\bibitem{richards2020decentralised}
D.~Richards, P.~Rebeschini, and L.~Rosasco, ``Decentralised learning with
  random features and distributed gradient descent,'' in \emph{Proc.\ 37th Int.
  Conf. Machine Learning (ICML)}.\hskip 1em plus 0.5em minus 0.4em\relax
  Virtual Conference: PMLR, Jul. 2020, pp. 8105--8115.

\bibitem{xu2019coke}
P.~Xu, Z.~Tian, Z.~Zhang, and Y.~Wang, ``Coke: Communication-censored kernel
  learning via random features,'' in \emph{2019 IEEE Data Science Workshop
  (DSW)}, Minneapolis, MN, USA, Jun. 2019, pp. 32--36.

\bibitem{shen2021distributed}
Y.~Shen, S.~Karimi-Bidhendi, and H.~Jafarkhani, ``Distributed and quantized
  online multi-kernel learning,'' \emph{IEEE Trans. Signal Process.}, vol.~69,
  pp. 5496--5511, Sep. 2021.

\bibitem{vapnik1992principles}
V.~Vapnik, ``Principles of risk minimization for learning theory,'' in
  \emph{Proc.\ 4th Int. Conf. Neural Information Processing Systems (NIPS)},
  Denver, Colorado, USA, Dec. 1991, pp. 831--838.

\bibitem{boyd2003subgradient}
S.~Boyd, L.~Xiao, and A.~Mutapcic, ``Subgradient methods,'' \emph{lecture notes
  of EE392o, Stanford University, Autumn Quarter}, 2003--2004.

\bibitem{ambrosio2008gradient}
L.~Ambrosio, N.~Gigli, and G.~Savar{\'e}, \emph{Gradient flows: in metric
  spaces and in the space of probability measures}.\hskip 1em plus 0.5em minus
  0.4em\relax Springer Science \& Business Media, 2008.

\bibitem{fiedler1973algebraic}
M.~Fiedler, ``Algebraic connectivity of graphs,'' \emph{Czechoslovak
  mathematical journal}, vol.~23, no.~2, pp. 298--305, 1973.

\bibitem{lecun1998gradient}
Y.~LeCun, L.~Bottou, Y.~Bengio, and P.~Haffner, ``Gradient-based learning
  applied to document recognition,'' \emph{Proc. IEEE}, vol.~86, no.~11, pp.
  2278--2324, Nov. 1998.

\bibitem{xiao2017fashion}
H.~Xiao, K.~Rasul, and R.~Vollgraf, ``Fashion-{MNIST}: a novel image dataset
  for benchmarking machine learning algorithms,'' \emph{arXiv preprint
  arXiv:1708.07747}, 2017.

\bibitem{krizhevsky2009learning}
A.~Krizhevsky and G.~Hinton, ``Learning multiple layers of features from tiny
  images,'' \emph{Technical report, University of Tronto}, 2009.

\bibitem{zhu2009complex}
H.~Zhu, H.~Luo, H.~Peng, L.~Li, and Q.~Luo, ``Complex networks-based
  energy-efficient evolution model for wireless sensor networks,'' \emph{Chaos,
  Solitons \& Fractals}, vol.~41, no.~4, pp. 1828--1835, Aug. 2009.

\bibitem{barabasi1999emergence}
A.-L. Barab{\'a}si and R.~Albert, ``Emergence of scaling in random networks,''
  \emph{Science}, vol. 286, no. 5439, pp. 509--512, Oct. 1999.

\bibitem{howard2019searching}
A.~Howard, M.~Sandler, G.~Chu, L.-C. Chen, B.~Chen, M.~Tan, W.~Wang, Y.~Zhu,
  R.~Pang, V.~Vasudevan \emph{et~al.}, ``Searching for {MobileNetV3},'' in
  \emph{Proc.\ IEEE/CVF Int. Conf. Computer Vision (ICCV)}, Seoul, Korea, Oct.
  2019, pp. 1314--1324.

\bibitem{he2016deep}
K.~He, X.~Zhang, S.~Ren, and J.~Sun, ``Deep residual learning for image
  recognition,'' in \emph{Proc.\ IEEE Conf. Computer Vision and Pattern
  Recognition (CVPR)}, Las Vegas, Nevada, USA, Jun. 2016, pp. 770--778.

\bibitem{ba2016layer}
J.~L. Ba, J.~R. Kiros, and G.~E. Hinton, ``Layer normalization,'' \emph{arXiv
  preprint arXiv:1607.06450}, Jul. 2016.

\bibitem{ioffe2015batch}
S.~Ioffe and C.~Szegedy, ``Batch normalization: Accelerating deep network
  training by reducing internal covariate shift,'' in \emph{Proc.\ 32nd Int.
  Conf. Machine Learning (ICML)}.\hskip 1em plus 0.5em minus 0.4em\relax Lille,
  France: PMLR, Jul. 2015, pp. 448--456.

\bibitem{sun2020improving}
H.~Sun, S.~Lu, and M.~Hong, ``Improving the sample and communication complexity
  for decentralized non-convex optimization: Joint gradient estimation and
  tracking,'' in \emph{Proc.\ 37th Int. Conf. Machine Learning (ICML)}, vol.
  119.\hskip 1em plus 0.5em minus 0.4em\relax Virtual Conference: PMLR, Jul.
  2020, pp. 9217--9228.

\bibitem{lu2020decentralized}
S.~Lu and C.~W. Wu, ``Decentralized stochastic non-convex optimization over
  weakly connected time-varying digraphs,'' in \emph{Proc.\ 45th Int. Conf.
  Acoustics, Speech and Signal Processing (ICASSP)}.\hskip 1em plus 0.5em minus
  0.4em\relax Virtual Conference: IEEE, May 2020, pp. 5770--5774.

\end{thebibliography}

%

\begin{IEEEbiography}[{\includegraphics[width=1in,height=1.25in,clip,keepaspectratio]{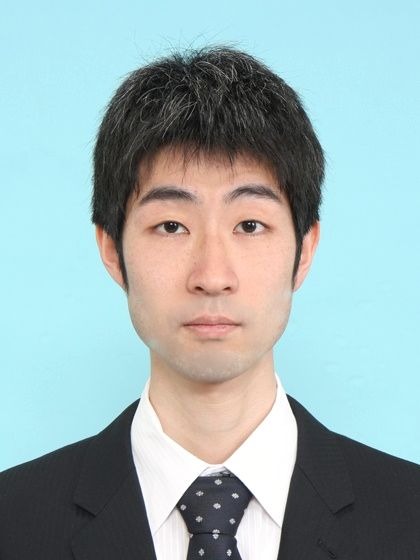}}]{Akihito Taya}
(S'12--M'17) received the B.E. degree in electrical and electronic engineering from Kyoto University, Kyoto, Japan in 2011,
and the master and Ph.D.\ degree in Informatics from Kyoto University in 2013 and 2019, respectively.
From 2013 to 2017, he joined Hitachi, Ltd., where he participated in the development of computer clusters.
He has been an assistant professor of the Aoyama Gakuin University, since 2019.
He received the IEEE VTS Japan Young Researcher's Encouragement Award and the IEICE Young Researcher's Award in 2012 and 2018, respectively.
His current research interests include distributed machine learning and human activity and emotion recognition using sensor networks.
He is a member of the IEEE.
\end{IEEEbiography}

\begin{IEEEbiography}[{\includegraphics[width=1in,height=1.25in,clip,keepaspectratio]{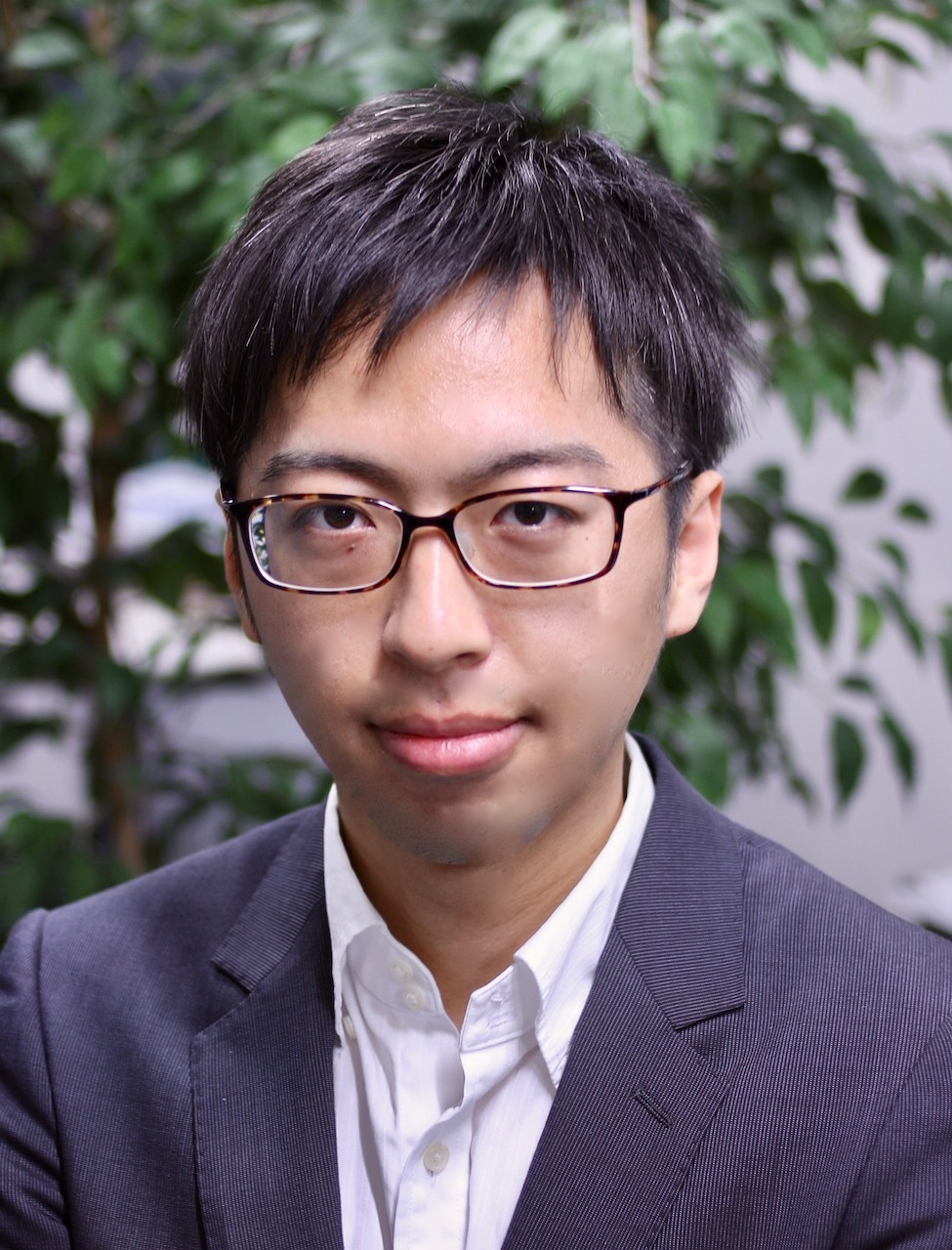}}]{Takayuki Nishio}
(S'11--M'14--SM'20) has been an associate professor in the School of Engineering, Tokyo Institute of Technology, Japan, since 2020. He received the B.E.degree in electrical and electronic engineering and the master's and Ph.D.\ degrees in informatics from Kyoto University in 2010, 2012, and 2013, respectively. He had been an assistant professor in the Graduate School of Informatics, Kyoto University from 2013 to 2020. From 2016 to 2017, he was a visiting researcher in Wireless Information Network Laboratory (WINLAB), Rutgers University, United States. His current research interests include machine learning-based network control, machine learning in wireless networks, and heterogeneous resource management.
\end{IEEEbiography}

\begin{IEEEbiography}[{\includegraphics[width=1in,height=1.25in,clip,keepaspectratio]{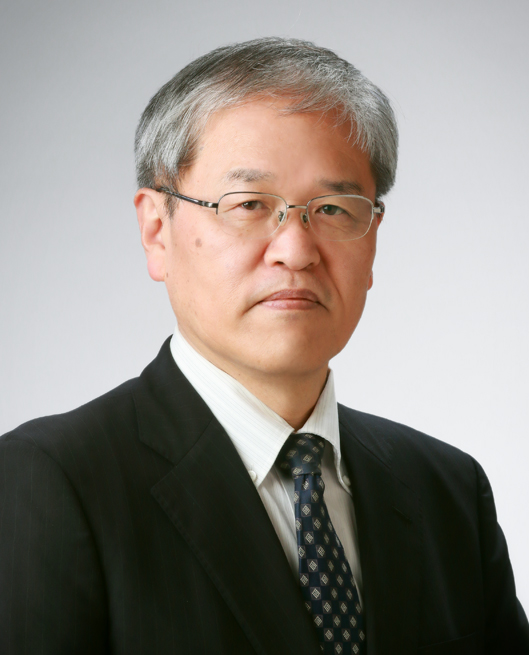}}]{Masahiro Morikura}
(M'82) received B.E., M.E. and Ph.D. degree in electronic engineering from Kyoto University, Kyoto, Japan in 1979, 1981 and 1991, respectively. He joined NTT in 1981, where he was engaged in the research and development of TDMA equipment for satellite communications.  From 1988 to 1989, he was with the communications Research Centre, Canada as a guest scientist. From 1997 to 2002, he was active in standardization of the IEEE802.11a based wireless LAN. He received Paper Award, Achievement Award and Distinguished Achievement and Contributions Award from the IEICE in 2000, 2006 and 2019, respectively. He also received Education, Culture, Sports, Science and Technology Minister Award in 2007 and Maejima Award from the Teishin association in 2008 and the Medal of Honor with Purple Ribbon from Japan’s Cabinet Office in 2015.
Dr. Morikura is now an emeritus professor of the Graduate School of Informatics, Kyoto University.  He is a Fellow of the IEICE and a member of IEEE.
\end{IEEEbiography}

\begin{IEEEbiography}[{\includegraphics[width=1in,height=1.25in,clip,keepaspectratio]{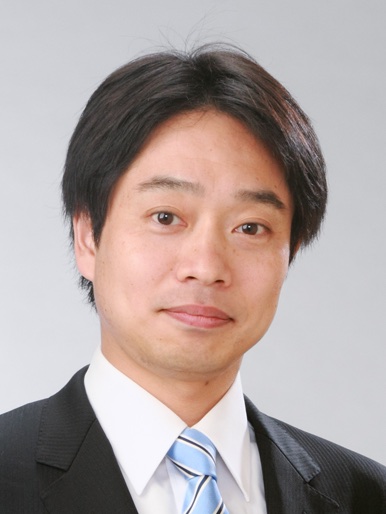}}]{Koji Yamamoto}
(S'03--M'06--SM'20) received the B.E. degree in electrical and electronic engineering from Kyoto University in 2002, and the master and Ph.D.\ degrees in Informatics from Kyoto University in 2004 and 2005, respectively.
From 2004 to 2005, he was a research fellow of the Japan Society for the Promotion of Science (JSPS).
Since 2005, he has been with the Graduate School of Informatics, Kyoto University, where he is currently an associate professor.
From 2008 to 2009, he was a visiting researcher at Wireless@KTH, Royal Institute of Technology (KTH) in Sweden.
He serves as an editor of IEEE Wireless Communications Letters, IEEE Open Journal of Vehicular Technology, and Journal of Communications and Information Networks, a symposium co-chair of GLOBECOM 2021, and a vice co-chair of IEEE ComSoc APB CCC.
He was a tutorial lecturer in ICC 2019.
His research interests include radio resource management, game theory, and machine learning.
He received the PIMRC 2004 Best Student Paper Award in 2004, the Ericsson Young Scientist Award in 2006.
He also received the Young Researcher's Award, the Paper Award, SUEMATSU-Yasuharu Award, Educational Service Award from the IEICE of Japan in 2008, 2011, 2016, and 2020, respectively, and IEEE Kansai Section GOLD Award in 2012.
He is a senior member of the IEEE and a member of the Operations Research Society of Japan.
\end{IEEEbiography}





\end{document}